\documentclass[11pt,a4paper]{article}

\usepackage{amsmath,amssymb,amsfonts}

\usepackage{jheppub-mod}

\newcommand{\eps}{\varepsilon}
\newcommand{\ph}{\varphi}

\newcommand{\mg}{\mathrm{mg}}
\newcommand{\el}{\mathrm{el}}

\DeclareMathOperator{\Det}{Det}

\newcommand{\tens}[1]{{\boldsymbol{#1}}}       
\newcommand{\ts}[1]{{\boldsymbol{#1}}}         

\newcommand{\grad}{{\tens{d}}}                 
\newcommand{\covd}{{\tens{\nabla}}}            
\newcommand{\cv}[1]{{\tens{\partial}}_{#1}}    
\newcommand{\pa}{\partial}                     

\newcommand{\A}[1]{A^{\!(#1)}}                 

\newcommand{\dg}{{N}}                            
\newcommand{\emp}{\mu}

\newcommand{\EMA}{\mathsf{A}}                     
\newcommand{\EMF}{\mathsf{F}}                     

\newcommand{\kop}{\mathcal{K}}
\newcommand{\lop}{\mathcal{L}}
\newcommand{\sop}{\mathcal{C}}
\newcommand{\mop}{\mathcal{J}}

\newcommand{\scs}{C}
\newcommand{\mcs}{m}

\newcommand{\efc}{\mathsf{Z}}

\newcommand{\env}[1]{{\tens{e}_{#1}}}        
\newcommand{\enf}[1]{{\tens{e}^{#1}}}        
\newcommand{\ehv}[1]{{\hat{\tens{e}}_{#1}}}  
\newcommand{\ehf}[1]{{\hat{\tens{e}}^{#1}}}  
\newcommand{\ezv}{{\hat{\tens{e}}_{0}}}      
\newcommand{\ezf}{{\hat{\tens{e}}^{0}}}      

\newcommand{\KT}[1]{\tens{k}_{(#1)}}                  
\newcommand{\KV}[1]{\tens{l}_{(#1)}}                  

\newcommand{\KTc}[1]{k_{(#1)}}                  

\newcommand{\xc}{\mathring{x}}

\newcommand{\Ac}[1]{\mathring{A}{}^{\!(#1)}}
\newcommand{\Uc}{\mathring{U}}

\newcommand{\be}{\begin{equation}}             
\newcommand{\ee}{\end{equation}}               

\title{Separation of Maxwell equations in Kerr--NUT--(A)dS spacetimes}

\author[a]{Pavel Krtou\v{s}}{%
\emailAdd{Pavel.Krtous@utf.mff.cuni.cz}
\affiliation[a]{%
Institute of Theoretical Physics,
Faculty of Mathematics and Physics, Charles University,\\
V~Hole\v{s}ovi\v{c}k\'ach~2, Prague, Czech Republic}

\author[b]{Valeri P. Frolov}
\emailAdd{vfrolov@ualberta.ca}
\affiliation[b]{%
Theoretical Physics Institute,
Department of Physics, University of Alberta,\\
Edmonton, Alberta, T6G 2G7, Canada}

\author[c]{David Kubiz\v{n}\'{a}k}{%
\emailAdd{dkubiznak@perimeterinstitute.ca}
\affiliation[c]{%
Perimeter Institute,
31 Caroline St. N. Waterloo Ontario, N2L 2Y5, Canada}

\abstract{
In this paper we explicitly demonstrate separability of the Maxwell equations in a wide class of higher-dimensional metrics which include the Kerr--NUT--(A)dS solution as a special case. Namely, we prove such separability for the most general  metric admitting the principal tensor (a non-degenerate closed conformal Killing--Yano 2-form). To this purpose we use a special ansatz for the electromagnetic potential, which we represent as  a product of a (rank 2) polarization tensor with the gradient of a potential function,  generalizing the ansatz recently proposed by Lunin. We show that for a special choice of the polarization tensor written in terms of the principal tensor, both the Lorenz gauge condition and the Maxwell equations  reduce to a composition of mutually commuting operators acting on the potential function. A solution to both these equations can be written in terms of an eigenfunction of these commuting operators. When incorporating a multiplicative separation ansatz, it turns out that the eigenvalue equations reduce to a set of separated ordinary differential equations with the eigenvalues playing a role of separability constants. The remaining ambiguity in the separated equations is related to an identification of $D-2$ polarizations of the electromagnetic field. We thus obtained a sufficiently rich set of solutions for the Maxwell equations in these spacetimes.
}

\arxivnumber{1803.02485}

\begin{document}
\maketitle

\newpage

\section{Introduction}
\label{sc:intro}

A method of separation of variables plays an important role in the theory of partial differential equations (PDEs). It allows one to reduce these equations to a set of ordinary differential equations (ODEs). The latter are simpler and can  be solved either analytically or by simple numerical methods. In particular, the separation of variables in the equations for physical fields in a curved space of a stationary black hole
allowed one to study many physical processes in the vicinity of these black holes such as propagation, scattering and capture of waves.  Separated equations for quasinormal modes were used to study the black hole stability and its ringing radiation. The method of separation of variables is also used to study the quantum Hawking effect.

Separation of variables in the physical field equations in the rotating black hole spacetime described by the Kerr geometry has a long history. It started in 1968, when Carter demonstrated that a scalar field equation can be solved by a method of separation of variables \cite{Carter:1968cmp}. In 1972, Teukolsky \cite{Teukolsky:1972,Teukolsky:1973} decoupled equations for the electromagnetic and gravitational perturbations  and  demonstrated that decoupled equations can be solved by the separation of variables. The massless neutrino equations were separated by Teukolsky \cite{Teukolsky:1973} and Unruh \cite{Unruh:1973} in 1973, and the  massive Dirac equations  were separated by Chandrasekhar \cite{Chandrasekhar:1976} and Page \cite{Page:1976} in 1976.

More recently, the development of brane-world models and the discussion of the possibility of mini black-hole creation in colliders attracted a lot of attention to the problem of separation of variables in higher-dimensional black hole spacetimes. This problem is rather straightforward for the (spherically symmetric) Tangherlini metric, which is a simple generalization of the Schwarzschild geometry. However, in the presence of rotation and NUT parameters it becomes quite complicated. One of the reasons is that even if the equations are separable in a given geometry, the separation occurs only in a very special coordinate system which is a priori not known.  Separation of variables in the  Klein--Gordon equation in the five-dimensional Myers--Perry metric was first demonstrated in \cite{FrolovStojkovic:2003a}, see also \cite{Kunduri:2005fq} for the 5-dimensional Kerr-(A)dS generalization. Page and collaborators \cite{Vasudevan:2005js,Vasudevan:2004mr,Vasudevan:2004ca} discovered that the Klein--Gordon equation is separable in a special case of the higher-dimensional Kerr-(A)dS spacetime, provided the black hole spin is restricted to two sets of equal rotation parameters. Upon this restriction, the explicit symmetry of the spacetime is enhanced and makes the separation of variables possible. Similar results, exploiting the enhanced symmetry of black holes arising from a restriction on rotation parameters, were obtained in \cite{Davis:2006hy,Chen:2006ui}.

The discovery \cite{Kubiznak:2006kt} of the principal tensor in the most general higher-dimensional Kerr--NUT--(A)dS spacetime \cite{Chen:2006xh} made it possible to solve the problem of separability for the Klein--Gordon equation without any restriction on rotation parameters \cite{Frolov:2006pe}. The principal tensor is a non-degenerate rank-2 closed conformal Killing--Yano tensor. The discussion of its remarkable properties can be found in a comprehensive review \cite{FrolovKrtousKubiznak:2017review}. This tensor generates a complete set (tower) of symmetries, which consists of Killing vectors and rank-2 Killing tensors. Moreover, the eigenvalues of the principal tensor,  together with the appropriate choice of the Killing coordinates, define special, the so called canonical coordinates. It is in these coordinates the separability property is valid for the Klein--Gordon equation. Later, it was shown that using the Killing vectors and Killing tensors in the Killing tower one can construct a full set of the corresponding first-order and second-order covariant differential operators, all mutually commuting, such that their eigenvalues coincide with the corresponding separation constants of the Klein--Gordon equation \cite{carter1987separability,Kolar:2015cha,Sergyeyev:2007gf,Frolov:2010cr}. This result demonstrated a close relationship between the separability structure of the spacetime and the existence of the principal tensor.

Let us write $D=2\dg+\eps$ for the number of spacetime dimensions, with $\eps=1$ for odd dimensions and $\eps=0$ for even ones. As shown in \cite{Houri:2007xz, Krtous:2008tb}, the most general metric that possesses the principal tensor admits $N$ arbitrary metric functions of one variable. We call such metrics off-shell. {For the on-shell metric, when the Einstein equations are imposed,} these metric functions reduce to polynomials, and, in the Lorentzian signature, we recover the \mbox{Kerr--NUT--(A)dS} solution \cite{Chen:2006xh}. Interestingly, the separation of variables in the Klein--Gordon equation remains valid for a general higher-dimensional off-shell geometry.

The separability of the massive Dirac equation in the higher-dimensional off-shell Kerr--NUT--(A)dS spacetimes was proved in \cite{OotaYasui:2008}, see also \cite{Cariglia:2011qb,Cariglia:2011yt,CarigliaEtal:2012b,Kubiznak:2010ig} for the intrinsic characterization of this separability in terms of the commuting operators. A partial success regarding the separation of variables for the special type of gravitational perturbations in these spacetimes was achieved in \cite{Kunduri:2006qa, Oota:2008uj}.

The question of separability of Maxwell equations in higher-dimensional rotating black hole spacetimes remained open for a long time. In four dimensions both electromagnetic strength field $\ts{\EMF}$ and its Hodge dual $*\ts{\EMF}$ are 2-forms. The complex self-dual and anti-self-dual 2-forms $\ts{\EMF}\,\pm\, i *\ts{\EMF}$ describe independent right- and left-polarization states of propagating electromagnetic waves. This property was essentially used in various  schemes of reduction of the Maxwell equations to a set of complex scalar equations and their further separation in the 4D Kerr--NUT--(A)dS metrics. Unfortunately, such a method cannot be generalized to higher dimensions.

A breakthrough in the problem of separability of the Maxwell equations in higher-dimensional rotating black hole spacetimes came in recent Lunin's paper \cite{Lunin:2017}. Lunin has proposed a special ansatz for the vector potential, which can be reformulated as
{${\EMA^a=B^{ab}\nabla_{\!b} Z}$,}
where $Z$ is a complex scalar function and ${\ts{B}}$ is a special tensor, which we call the polarization tensor. In his work Lunin has written down the ansatz for the vector potential in a special frame, effectively specifying the polarization tensor. He used special coordinates, different from the Myers--Perry coordinates, which are closely related to the canonical coordinated connected with the principal tensor \cite{Frolov:2008jr, Chervonyi:2015ima}. In this setting Lunin demonstrated \cite{Lunin:2017} that the Maxwell equations in the higher-dimensional Kerr--(A)dS spacetimes imply separable equations for the function $Z$.

In this paper we propose an essential development of Lunin's approach. Our analysis is performed for general off-shell metrics which admit a non-degenerate principal tensor~$\ts{h}$. We first find a covariant (coordinate-independent) expression for the polarization tensor~$\ts{B}$ in terms of the principal tensor. Next we show that the Lorenz condition for the vector potential, which becomes a second-order wave-like operator acting on the function ${Z}$, can be understood as a composition  of ${\dg}$ second-order commuting operators. These, supplemented with derivatives along the explicit spacetime symmetries, form a system of ${D}$ mutually commuting operators which possess a common system of eigenfunctions. Among these eigenfunctions one can find those solving the Lorenz condition. It turns out that these solutions are labeled by ${D}$ eigenvalues and by a discrete choice of ${\dg-1}$ (complex) polarizations.

Moreover, we show that the Maxwell equations can be reduced to the simultaneous validity of the Lorenz condition and of another wave-like equation. This additional equation is, surprisingly, also a composition  of the same operators which have been identified in the Lorenz condition. Solution of the Maxwell equations can thus be constructed from the eigenfunctions of these operators, provided that one of the eigenvalues is set to zero---the condition reflecting the massless character of the electromagnetic field.

Next we arrive at the key observation, namely that the structure of commuting operators allows one to find a system of eigenfunctions by the method of separation of variables. Each eigenfunction can be written as a product of functions of one variable. The Lorenz condition, as well as the Maxwell equations, then require that a composition of conditions for these functions should vanish. Since these conditions depend on different variables, they must vanish independently, with a freedom of a choice of separation constants. These separation constants turn out to be exactly the eigenvalues of the eigenfunctions under investigation.

We thus demonstrate that a solution of the Maxwell equations can be found using a multiplicative separation ansatz and reduces to a solution of ${\dg}$ second-order ordinary differential equations. Such solutions are labeled by ${D-1}$ separation constants.

When constructing separable solutions we identify independent polarizations associated with each choice of the separation constants. This identification is done in a different manner than in the work of Lunin. Namely, we find ${D-3}$ generic polarizations, we call them the magnetic polarizations, and one special stationary one, which we call the electric polarization. We discuss also different parametrizations of the space of solutions which may be closer to the method of Lunin.

Let us emphasize that our  construction of the separable solutions of the Maxwell equations (i) is valid for an arbitrary off-shell metric admitting the principal tensor, and (ii) the proof of the separability is done in a totally analytic way and presented in the paper with all necessary details.

The paper is organized as follows. The properties of the principal tensor and of the \mbox{off-shell} Kerr--NUT--(A)dS metrics are reviewed in section~\ref{sc:geometry}. In section~\ref{sc:FieldAnsatz} we introduce the ansatz for the polarization tensor and discuss its properties. The Lorenz condition and the form of the Maxwell equations for this ansatz are derived in section \ref{sc:FieldEqs}. The commuting operators with a common set of eigenfunctions are introduced in section \ref{sc:StrcEqs}. Sections \ref{sc:MagPol} and \ref{sc:ElPol} contain a discussion of special types of solutions of the higher-dimensional Maxwell equations. The separable character of these solutions and the meaning of the separation constants are a subject of section~\ref{sc:SepVar}. Section~\ref{sc:aligned} is devoted to a discussion of special aligned fields previously studied in the literature \cite{Krtous:2007,Chen:2007fs,Kolar:2015cha}. Section~\ref{sc:summary} contains the summary of obtained results, as well as a discussion of some unsolved problems. In this paper we extensively use the material and notations of review~\cite{FrolovKrtousKubiznak:2017review}.

\section{Off-shell Kerr--NUT--(A)dS geometry}
\label{sc:geometry}

This section contains a brief summary of properties  of spacetimes admitting the  principal tensor.
A thorough discussion of these spacetimes, the principal tensor, the associated Killing tower, and the on-shell and off-shell Kerr--NUT--(A)dS geometries can be found in the recent review~\cite{FrolovKrtousKubiznak:2017review}.

\subsection{Principal tensor and metric}

In what follows we denote by $D=2\dg+\eps$ the number of dimensions.
We are interested in spaces which possess the principal tensor: a non-degenerate closed conformal Killing--Yano 2-form. The principal tensor $\ts{h}$ satisfies the following equation:
\begin{equation}
    \nabla_{\!c} h_{ab} = g_{ca}\xi_b - g_{cb}\xi_a\;,\label{PrincTdef}
\end{equation}
where $\ts{\xi}$ is a primary Killing vector,
\begin{equation}\label{PrimaryVec}
    \xi_a = \frac1{D-1}\nabla^b h_{ba}\;.
\end{equation}
The non-degeneracy of the principal tensor essentially means that the principal tensor has non-degenerate imaginary eigenvalues $\pm i x_\mu$, $(\mu=1,\ldots,\dg)$, and that $x_\mu$ are independent functions which, when supplemented with an appropriate set of Killing angles, can be used as {\em canonical coordinates}. The metric can be written in a formally Euclidian Darboux frame in which the principal tensor has a semi-diagonal form:
\begin{gather}
   \ts{g} = \sum_\mu \bigl( \enf{\mu} \enf{\mu} + \ehf{\mu} \ehf{\mu} \bigr)
      + \eps\, \ezf \ezf\;, \label{mtrcfr}\\
   \ts{h} = \sum_\mu x_\mu \, \enf{\mu} \wedge \ehf{\mu} \;. \label{hfr}
\end{gather}

In the canonical coordinates, the metric reads\footnote{%
We write sums over coordinate indices $\mu,\nu,\dots$ and $k,l,\dots$ explicitly, but we usually do not indicate their ranges. If they are not indicated, we assume $\sum_\mu=\sum_{\mu=1}^{\dg}$ and $\sum_k=\sum_{k=0}^{\dg-1}$.}
\begin{equation}\label{metric}
\ts{g}
  =\sum_{\mu=1}^\dg\;\biggl[\; \frac{U_\mu}{X_\mu}\,{\grad x_{\mu}^{2}}
  +\, \frac{X_\mu}{U_\mu}\,\Bigl(\,\sum_{j=0}^{\dg-1} \A{j}_{\mu}\grad\psi_j \Bigr)^{\!2}
  \;\biggr]
  +\eps\frac{c}{\A{\dg}}\Bigl(\sum_{k=0}^\dg \A{k}\grad\psi_k\!\Bigr)^{\!2}\;,
\end{equation}
where $\A{k}$, $\A{j}_\mu$, and $U_\mu$ are explicit polynomial functions of coordinates $x_\nu^2$,
\begin{equation}\label{AUdefs}
\begin{gathered}
  \A{k}=\!\!\!\!\!\sum_{\substack{\nu_1,\dots,\nu_k\\\nu_1<\dots<\nu_k}}\!\!\!\!\!x^2_{\nu_1}\dots x^2_{\nu_k}\;,
\quad
  \A{j}_{\mu}=\!\!\!\!\!\sum_{\substack{\nu_1,\dots,\nu_j\\\nu_1<\dots<\nu_j\\\nu_i\ne\mu}}\!\!\!\!\!x^2_{\nu_1}\dots x^2_{\nu_j}\;,\\
  U_{\mu}=\prod_{\substack{\nu\\\nu\ne\mu}}(x_{\nu}^2-x_{\mu}^2)\;,
\end{gathered}
\end{equation}
and each metric function ${X_\mu}$ is a function of a single coordinate ${x_\mu}$:
\begin{equation}\label{Xfcdependence}
    X_\mu=X_\mu(x_\mu)\;.
\end{equation}
If these functions are chosen arbitrary, the geometry in general does not satisfy the Einstein equations and we call it off-shell. If the Einstein equations are imposed, $X_\mu$~must take a form of specific polynomials
\cite{Chen:2006xh,Houri:2007xz}:
\begin{equation}\label{Xsol}
  X_\mu =
  \begin{cases}
    {\displaystyle -2b_\mu\, x_\mu + \sum_{k=0}^{\dg}\, c_{k}\, x_\mu^{2k}}
      \qquad &\text{for $D$ even}\;,\\
    {\displaystyle -\frac{c}{x_\mu^2} - 2b_\mu + \sum_{k=1}^{\dg}\, c_{k}\, x_\mu^{2k}}
      \qquad &\text{for $D$ odd}\;.
  \end{cases}
\end{equation}
Here the parameter $c_\dg$ gives the cosmological constant, while other parameters are related to the mass, NUT parameters, and rotations, see \cite{FrolovKrtousKubiznak:2017review} for more details. In particular, in the Lorentzian signature we would recover the on-shell Kerr--NUT--(A)dS spacetimes \cite{Chen:2006xh}. However, in what follows we do not assume this specific choice and the subsequent discussion
is valid for the full off-shell family of spacetimes.

The Darboux frame of 1-forms ${\enf\mu,\,\ehf\mu}$ (${\mu=1,\dots,\dg}$), and ${\ezf}$ (in odd dimensions) read \begin{equation}\label{Darbouxformfr}
\enf\mu = {\Bigl(\frac{U_\mu}{X_\mu}\Bigr)^{\!\frac12}}\grad x_{\mu}\;,\quad
\ehf\mu = {\Bigl(\frac{X_\mu}{U_\mu}\Bigr)^{\!\frac12}}
  \sum_{j=0}^{\dg-1}\A{j}_{\mu}\grad\psi_j\;,\quad
\ezf = {\Bigl(\frac{c}{\A{\dg}}\Bigr)^{\frac12}}\,\sum_{k=0}^{\dg}\A{k}\grad\psi_k\,,
\end{equation}
with the dual vector frame ${\env\mu,\,\ehv\mu}$  and ${\ezv}$ given by
\begin{equation}\label{Darbouxvecfr}
\env\mu = {\Bigl(\frac{X_\mu}{U_\mu}\Bigr)^{\!\!\frac12}}{\cv{x_\mu}}\,,\quad
\ehv\mu = {\Bigl(\frac{U_\mu}{X_\mu}\Bigr)^{\!\!\frac12}}\!\sum_{k=0}^{\dg{-}1{+}\eps}\!
  {\frac{(-x_{\mu}^2)^{\dg{-}1{-}k}}{U_{\mu}}}\,{\cv{\psi_{k}}}\,,\quad
\ezv= \bigl(c \A{\dg}\bigr)^{\!-\frac12}{\cv{\psi_{\dg}}}\,.
\end{equation}
The primary Killing vector in the canonical coordinates and the Darboux frame is
\begin{equation}
   \ts{\xi} =  \cv{\psi_0} = \sum_\mu \Bigl(\frac{X_\mu}{U_\mu}\Bigr)^{\!\frac12}\ehv{\mu}
       +\eps \Bigl(\frac{c}{\A{\dg}}\Bigr)^{\!\frac12}\ezv\label{xifr}\;.
\end{equation}
The square of the principal tensor is a conformal Killing tensor,
\begin{equation}\label{Qdef}
  Q_{ab} = h_{ac} h_{bd} g^{cd}\,,
\end{equation}
which identifies ${\env{\mu}}$ and ${\ehv{\mu}}$ as its eigenvectors with the eigenvalue $x_\mu^2$:
\begin{equation}
   \ts{Q} = \sum_\mu x_\mu^2 \bigl( \enf{\mu} \enf{\mu} + \ehf{\mu} \ehf{\mu} \bigr)\;. \label{Qfr}
\end{equation}

The metric \eqref{metric} describes a wide class of geometries, both Riemannian and Lorentzian, subject to possible Wick rotations of coordinates and a choice of signs of metric functions. We will not attempt to classify this family of geometries here, see \cite{FrolovKrtousKubiznak:2017review} for a discussion. We just recall that the family contains the on-shell Kerr--NUT--(A)dS black holes, which, when the NUT parameters are turned off, are equivalent to the Myers--Perry spacetimes \cite{MyersPerry:1986} with possibly a  cosmological constant \cite{Gibbons:2004uw, Gibbons:2004js}. The coordinates used here generalize Carter's coordinates known in four dimension, with ${x_\dg}$ being the Wick rotated radial coordinate and other ${x_\nu}$ corresponding to (cosine of) latitudinal angular coordinates.
Killing coordinates ${\psi_k}$ correspond to explicit symmetries of the space: time and longitudinal angles. However, this relation is not direct, see appendix~\ref{sc:KillingAngles} for more details.

\subsection{Killing tower}

The principal tensor guarantees the existence of a rich symmetry structure, the so called Killing tower of Killing and Killing--Yano objects \cite{Krtous:2006qy}. Here we are going to introduce only the Killing tensors and Killing vectors since they are directly related to the symmetries of various fields in the studied spaces. The Killing tower can be defined in terms of generating functions. First we define a $\beta$-dependent conformal Killing tensor $\ts{q}(\beta)$,
\begin{equation}\label{qdef}
    \ts{q}(\beta) = \ts{g} + \beta^2 \ts{Q}\;,
\end{equation}
and scalar functions $A(\beta)$ and $A_\mu(\beta)$,
\begin{align}
  A(\beta) &= \sqrt{\frac{\Det\ts{q}(\beta)}{\Det{\ts{g}}}}
    = \prod_\nu (1+\beta^2 x_\nu^2)
    \;,\label{Adef}\\
  A_\mu(\beta) &= \frac{A(\beta)}{1+\beta^2x_\mu^2}
    = \prod_{\substack{\nu\\\nu\neq\mu}} (1+\beta^2 x_\nu^2)
    \;.\label{Amudef}
\end{align}
In the following, we usually skip the argument $\beta$ to keep the expressions more compact.

The generating Killing tensor $\ts{k}(\beta)$ and the generating Killing vector $\ts{l}(\beta)$ are defined as
\begin{align}
    \ts{k} &= A\, \ts{q}^{-1}\;,\label{kdef}\\
    \ts{l} &= \ts{k}\cdot\ts{\xi}\;.\label{ldef}
\end{align}
The Killing tower of Killing tensors $\KT{j}$ and Killing vectors $\KV{j}$ is given by an expansion in $\beta$
\begin{gather}
    \ts{k}(\beta) = \sum_{j}\KT{j}\, \beta^{2j}\;,\label{KTfcexp}\\
    \ts{l}(\beta) = \sum_{j}\KV{j}\,\beta^{2j}\;.\label{KVfcexp}
\end{gather}
Note that only terms for $j=0,1,\dots,\dg{-}1{+}\eps$ are nonvanishing.
One also has
\begin{equation}\label{Asexp}
    A(\beta)  = \sum_{j=0}^\dg \A{j}\, \beta^{2j}\; ,\quad
    A_\mu(\beta)  = \sum_{j=0}^{\dg-1}\A{j}_\mu\,\beta^{2j}\; ,
\end{equation}
with $\A{j}$, $\A{j}_\mu$ being the standard symmetric polynomials introduced in \eqref{AUdefs}.

In the Darboux frame, the generating Killing tensor and Killing vector are
\begin{gather}
  \ts{k} = \sum_\mu A_\mu\,\bigl(\env{\mu}\env{\mu}+\ehv{\mu}\ehv{\mu}\bigr)
      +\eps A\,\ezv\ezv\;,\label{kfr}\\
  \ts{l} = \sum_\mu A_\mu \Bigl(\frac{X_\mu}{U_\mu}\Bigr)^{\!\frac12}\ehv{\mu}
       +\eps A\,\Bigl(\frac{c}{\A{\dg}}\Bigr)^{\!\frac12}\ezv\;,
\end{gather}
while in coordinates they read
\begin{gather}
  \ts{k}
  =\sum_{\mu}\; \frac{A_\mu}{U_\mu}\biggl[\; X_\mu\,{\cv{x_{\mu}}^2}
  + \frac{1}{X_\mu}\,\Bigl(\,\sum_{k=0}^{\dg-1+\eps}\!
    {(-x_{\mu}^2)^{\dg-1-k}}\,\cv{\psi_k}\Bigr)^{\!2}\;\biggr]
  +\eps\,\frac{A}{\A{\dg}}\cv{\psi_n}^2
  \;,\label{kcoor}\\[1ex]
  \ts{l} = \sum_{j=0}^{\dg-1+\eps} \beta^{2j}\cv{\psi_j}\,.  \label{lcoor}
\end{gather}
Similar expressions for individual Killing tensors and Killing vectors from the tower are obtained by a simple ${\beta}$-expansion. We emphasize only
\begin{equation}\label{KVjcoor}
  \KV{j} = \cv{\psi_j}\;.
\end{equation}

A trace $k^a{}_a(\beta)$ of the generating Killing tensor is
\begin{equation}\label{trkdef}
    k^a{}_a = 2 \sum_\mu A_\mu + \eps A = -\beta\frac{d}{d\beta}(\beta^{-D} A)\;,
\end{equation}
since the traces of individual Killing tensors are
\begin{equation}\label{trkj}
    \KTc{j}{}^a{}_a = 2 \sum_\mu \A{j}_\mu + \eps \A{j} = (D-2j)\A{j}\;,
\end{equation}
and
\begin{equation}\label{KTtrexp}
   k^a{}_a(\beta) = \sum_{j=0}^{\dg-1}\KTc{j}{}^a{}_a\, \beta^{2j}\;.
\end{equation}

The covariant derivative of the generating Killing tensor is \cite{FrolovKrtousKubiznak:2017review}
\begin{equation}\label{covdKTb}
  \nabla^{c} k^{ab} =
    \frac{2\beta^2}{A}\bigl(
    k^{ab}\,k^{cn}\,h_{n}{}^{m}+h^{m}{}_{n}\,k^{n(a}\,k^{b)c}
    +k^{m(a}\,k^{b)n}\,h_{n}{}^{c}\bigr)\,\xi_m\,.
\end{equation}
The contraction gives
\begin{equation}\label{kdiv}
    \nabla_{\!n}k^{na}
     = \frac{\beta^2}{A}
       (2 k{}^a{}_m k{}^m{}_n h{}^n{}_l\xi^l - k^c{}_c\, k{}^a{}_m h{}^m{}_n \xi^n)
     = \Bigl(k^{an} - \frac12 k^c{}_c\, g^{an}\Bigr)\,\frac1A\nabla_{\!n} A\;,
\end{equation}
where we used another useful relation:
\begin{equation}\label{dA}
  \frac12\nabla_{\!a} A = \beta^2\, h_{an}\, l^n\,.
\end{equation}

Finally, the generating Killing tensor commutes with the principal tensor in the sense of matrix multiplication
\begin{equation}\label{khcom}
    h{}^a{}_n k{}^n{}_b = k{}^a{}_n h{}^n{}_b\;.
\end{equation}

All these definitions and relations have been discussed in the literature and are reviewed in \cite{FrolovKrtousKubiznak:2017review}.

\section{Field ansatz}
\label{sc:FieldAnsatz}

We want to study a test electromagnetic field in the background of the off-shell Kerr--NUT--(A)dS spacetime. We are looking for a field which solves the Maxwell equations in a separable form. However, we have to face the fact that the electromagnetic field has several components and that these components are coupled together. The long-standing problem of decoupling the Maxwell equations in higher dimension was successfully attacked  by Lunin \cite{Lunin:2017} in the case of the field in the background of the Myers--Perry and Kerr--(A)dS black holes.

In four dimension we have demonstrated \cite{FrolovKrtousKubiznak:2018a} that Lunin's ansatz for the field can be reformulated covariantly in terms of the principal tensor. Similarly, in higher dimensions we assume that the electromagnetic vector potential\footnote{%
Unfortunately, the letter $A$ is heavily used for various alternatives of metric functions, namely, $A$, $A_\mu$, $\A{k}$, $\A{k}_\mu$. Therefore we use Serif font for the vector potential $\EMA_a$ to avoid a confusion. Consistently, we use $\EMF_{ab}$ for the field strength.}
$\ts{\EMA}$ has the form
\begin{equation}\label{Aansatz}
    \EMA^a = B^{ab} \nabla_{\!b}Z\;.
\end{equation}
Here, $Z$ is an auxiliary complex scalar function that plays a role of a kind of scalar potential for the  vector potential $\ts{\EMA}$. This function will be searched for and found in a multiplicative separated form.

Let us first concentrate on the polarization tensor $\ts{B}$ in the ansatz \eqref{Aansatz}. It is defined in terms of the principal tensor $\ts{h}$ as
\begin{equation}\label{Bdef}
    B^{ac} (g_{cb}-\beta h_{cb}) = \delta^a_b\;.
\end{equation}
$\ts{B}(\beta)$ thus depends on a parameter $\beta$, which is in general complex.

Since \eqref{Bdef} means that $\ts{B}=(\ts{g}-\beta\ts{h})^{-1}$, the `symmetric square' of $\ts{B}$ is closely related to the generating Killing tensor $\ts{k}$\footnote{Here and later we use a dot to denote a contraction of two subsequent tensors with respect to their two neighbor indices. For example, for two tensors with components $X_{ab}$ and $Y^{cd}$, $\ts{X}\cdot \ts{Y}$ means a tensor with components $X_{ac}Y^{cb}$.}
\begin{equation}\label{BBisAKder}
    \ts{B}\cdot\ts{g}\cdot\ts{B}^T
    = (\ts{g}-\beta\ts{h})^{\!-1}\cdot\ts{g}\cdot(\ts{g}+\beta\ts{h})^{\!-1}
    = (\ts{g}+\beta^2\ts{Q})^{\!-1}
    = \frac1A\,\ts{k}\,,
\end{equation}
or in indices,
\begin{equation}\label{BBisAK}
    B^{ak}B^{bl}g_{kl} = \frac1A k^{ab}\;.
\end{equation}
Inverting $\ts{B}^T$, we find
\begin{equation}\label{Bkrel}
    B^{ab} = \frac1A\,g^{am}(g_{mn}+\beta h_{mn})k^{nb}\;.
\end{equation}
From here we can read off the symmetric and antisymmetric parts of $\ts{B}$:
\begin{equation}\label{Bevenodd}
\begin{aligned}
    B^{(ab)} &= \frac1A\,k^{ab}\;,\\
    B^{[ab]} &= \frac{\beta}{A}\,h^a{}_n k^{nb}= \frac{\beta}{A}\,k^{an} h_n{}^{b}\;.
\end{aligned}
\end{equation}
Thanks to this, the trace of $\ts{B}$ is
\begin{equation}\label{Btr}
    B{}^n{}_n = \frac{k^a{}_a}{A}\;.
\end{equation}

Taking a covariant derivative of definition \eqref{Bdef} and employing relation \eqref{PrincTdef} and \eqref{Bdef}, one finds
\begin{equation}\label{Bder}
    \nabla_{\!c}B^{ab} = \beta\,(B^a{}_c\, \xi_n B^{nb}-B^{an}\xi_n\, B_c{}^b)\;.
\end{equation}
Contractions yield
\begin{equation}\label{Bdiv}
\begin{aligned}
    \nabla_{\!n}B^{nb} &=  \frac{\beta}{A} ( k^a{}_a \xi_n B^{nb} - \xi_n k^{nb} )\;,\\
    \nabla_{\!n}B^{an} &=  \frac{\beta}{A} ( \xi_n k^{na} - k^b{}_b B^{an} \xi_n )\;.
\end{aligned}
\end{equation}

\section{Field equations}
\label{sc:FieldEqs}

We use the ansatz \eqref{Aansatz} to obtain solutions of the Maxwell equations in the higher dimensional off-shell Kerr-NUT-(A)dS spacetimes. We proceed as follows. First, we impose the Lorenz condition on the potential $\ts{\EMA}$ and demonstrate that the obtained second order equation for the potential $Z$ allows the separation of variables in the canonical coordinates. After this we show that the Maxwell field equations are satisfied provided (i) the Lorenz equation is valid and (ii) an additional equation for $Z$ is valid. And finally, we show that this additional equation is also satisfied provided $Z$ obeys the separable equation, obtained from the Lorenz condition.

For simplicity, starting with this section we restrict ourselves to even dimensions. Thanks to that the coordinate expressions for differential operators are slightly shorter. The full expression for scalar operators in odd dimensions can be found in \cite{Sergyeyev:2007gf}. Similar expressions could be written for the electromagnetic case.

\subsection{Covariant form of the Lorenz condition}

Let us start investigating the Lorenz condition
\begin{equation}\label{LorCond}
   \nabla_{\!a}\EMA^a = 0\;.
\end{equation}
In the appendix (see \eqref{proofnablaBnablaZ1}, \eqref{proofnablaBnablaZ2}) we show that the divergence of the vector potential \eqref{Aansatz} reads
\begin{equation}\label{nablaBnablaZconf}
    \nabla_{\!m}\EMA^m = \nabla_{\!m}\bigl(B^{mn}\nabla_{\!n}Z\bigr)
    = \nabla_{\!m}\Bigl(\frac1{A}k^{mn}\nabla_{\!n} Z\Bigr)
          +\frac\beta{A}\Bigl(\frac{k^a{}_a}{A}-1\Bigr)\,l^n\nabla_{\!n} Z\;.
\end{equation}
Taking the factor $1/A$ out in the first term, one can also write
\begin{equation}\label{nablaBnablaZ}
    \nabla_{\!m}\bigl(B^{mn}\nabla_{\!n}Z\bigr)
    = \frac1{A}\,\nabla_{\!m}\bigl(k^{mn}\nabla_{\!n} Z\bigr)
    +\frac1{A}\,\biggl(-\frac1{A}(\nabla_{\!m}A)k^{mn}
          +\beta\Bigl(\frac{k^a{}_a}{A}-1\Bigr)l^n\biggr)\nabla_{\!n} Z\;.
\end{equation}

\subsection{Coordinate form of the Lorenz condition}

The first term in \eqref{nablaBnablaZ} is, up to a prefactor $1/A$, the scalar wave operator associated with the Killing tensor $\ts{k}$. Such operators have been studied in \cite{Sergyeyev:2007gf} and we can use its coordinate form \eqref{kopcoor} reviewed in the appendix~\ref{sc:scWaveSep}.
Using \eqref{Adef} and \eqref{kcoor}, we find that the first term in the brackets in \eqref{nablaBnablaZ}, which is linear in $\covd Z$, has the form
\begin{equation}\label{term1}
    -\frac1{A}(\nabla_{\!m}A)k^{mn} \nabla_{\!n} Z
    =- \sum_\nu \frac{A_\nu}{U_\nu} X_\nu
    \frac{2\beta^2 x_\nu}{1+\beta^2 x_\nu^2}\frac{\pa}{\pa x_\nu} Z\;.
\end{equation}
In the appendix we prove the identity \eqref{ka1idproof2},
\begin{equation}\label{ka1id}
    \frac{k^a{}_a}{A}-1 = \beta^{2-D} \sum_\nu \frac{A_\nu}{U_\nu}
      \frac{1-\beta^2x_\nu^2}{1+\beta^2x_\nu^2}\;,
\end{equation}
which allows us to express the second term in the brackets linear in $\covd Z$ in \eqref{nablaBnablaZ},
\begin{equation}\label{term2}
    \beta\Bigl(\frac{k^a{}_a}{A}-1\Bigr)\, l^n \nabla_{\!n} Z
    = \beta\sum_\nu \frac{A_\nu}{U_\nu}
      \frac{1-\beta^2x_\nu^2}{1+\beta^2x_\nu^2}\;
      \beta^{2(1-\dg)}\sum_j\beta^{2j}\frac{\pa}{\pa\psi_j}Z\,.
\end{equation}
Putting these together, the coordinate expression for the divergence of the vector potential \eqref{nablaBnablaZ} reads
\be \label{nablaBnablaZcoor}
    \nabla_{\!m}\bigl(B^{mn}\nabla_{\!n}Z\bigr)
    =\frac1{A}\sum_\nu \frac{A_\nu}{U_\nu}\tilde{\sop}_\nu Z\;,
\ee
where
\begin{align}\label{sun33}
\tilde{\sop}_\nu
    =&(1{+}\beta^2x_\nu^2)\frac{\pa}{\pa x_\nu}
      \Bigl[\frac{X_\nu}{1{+}\beta^2x_\nu^2}\frac{\pa}{\pa x_\nu}\Bigr]
    + \frac1{X_\nu} \Bigl[\sum_j (-x_\nu^2)^{\dg{-}1{-}j}\frac{\pa}{\pa\psi_j}\Bigr]^2\nonumber\\
    &       +\beta\frac{1{-}\beta^2x_\nu^2}{1{+}\beta^2x_\nu^2}\;
      \beta^{2(1-\dg)}\sum_j\beta^{2j}\frac{\pa}{\pa\psi_j}\,.
\end{align}

\subsection{Covariant form of the Maxwell equations}

The left-hand side of the Maxwell equations written in terms of the vector potential reads
\begin{equation}\label{MaxwellA}
    \nabla_{\!n}\EMF^{an} = - \Box \EMA^a
      + R^a{}_n \EMA^n
      + \nabla^a\bigl(\nabla_{\!n}\EMA^n\bigr)
      \;,
\end{equation}
with $\Box \equiv \nabla_{\!m}\nabla^{m}$. Inserting ansatz \eqref{Aansatz}, we get
\begin{equation}\label{MaxwellZpre}
    \nabla_{\!n}\EMF^{an} =
      - \nabla_{\!m}\nabla^m (B^{an}\nabla_{\!n}Z) + R^a{}_m B^{mn}\nabla_{\!n}Z
     + \nabla^a\bigl(\nabla_{\!m}(B^{mn}\nabla_{\!n}Z)\bigr)
      \;.
\end{equation}
In appendix~\ref{sc:Proofs} we derive a nontrivial identity \eqref{boxA} for the first two terms, which gives us
\begin{equation}\label{MaxwellZ}
 \begin{split}
    \nabla_{\!n}\EMF^{an} =
    &-B^{am}\nabla_{\!m}\bigl(\Box Z
    +2\beta \xi_k B^{kn}\nabla_{\!n}Z\bigr)\\
    &+2\beta B^{ak}\xi_k \nabla_{\!m}(B^{mn}\nabla_{\!n}Z)
    +\nabla^a\bigl(\nabla_{\!m}(B^{mn}\nabla_{\!n}Z)\bigr)
    \;.
 \end{split}
\end{equation}
Clearly, if the Lorenz condition is satisfied, $\nabla_{\!m}\bigl(B^{mn}\nabla_{\!n}Z\bigr)=0$, then the last two terms vanish and the vacuum Maxwell equations read
\begin{equation}\label{MaxwellZvac}
    B^{am}\nabla_{\!m}\Bigl(\Box Z
    +2\beta \xi_k B^{kn}\nabla_{\!n}Z\Bigr) = 0
    \;.
\end{equation}

\subsection{Coordinate form of the Maxwell equations}

We already know the coordinate form of the Lorenz condition, so we concentrate on the operator
\begin{equation}\label{BoxModOp}
    \bigl(\Box +2\beta \xi_k B^{kn}\nabla_{\!n}\bigr)Z\;.
\end{equation}
The box operator is given by expression for ${\kop_0}$ in \eqref{kopjcoor}. The second term in the bracket, using \eqref{Bkrel}, \eqref{ldef}, and \eqref{dA}, yields
\begin{equation}\label{xiBdZ1}
  2\beta\xi_k B^{kn}\nabla_{\!n}Z
    =\beta\frac{2}{A}l^n\nabla_{\!n} Z - \frac1{A}(\nabla^n A)\nabla_{\!n}Z\;.
\end{equation}
Employing identity \eqref{1overAid2} and the coordinate form \eqref{lcoor} in the first term and
$\frac1A\covd A = \covd\log A = \sum_\nu \covd\log(1{+}\beta^2x_\nu^2)$ with the index raised using the coordinate metric component $g^{\nu\nu}=\frac{X_\nu}{U_\nu}$ in the second term, we obtain
\begin{equation}\label{xiBdZ2}
  2\beta\xi_k B^{kn}\nabla_{\!n}Z
    =-\sum_\nu\frac{X_\nu}{U_\nu}\frac{2\beta^2x_\nu}{1{+}\beta^2x_\nu^2} \frac{\pa}{\pa x_\nu}Z
    +\beta\sum_\nu\frac1{U_\nu} \frac{1{-}\beta^2x_\nu^2}{1{+}\beta^2x_\nu^2}\beta^{2(1{-}\dg)}
    \sum_{k}\beta^{2k}\frac{\pa}{\pa\psi_k}Z
    \;.
\end{equation}
The first term nicely combines with the coordinate expression for the box, yielding
\begin{equation}\label{BoxModOpcoor}
    \bigl[\Box +2\beta \xi_k B^{kn}\nabla_{\!n}\bigr]Z
    =\sum_\nu \frac{1}{U_\nu}\tilde{\sop}_\nu Z
\end{equation}
for the operator \eqref{BoxModOp}, with $\tilde{\sop}_\nu$ defined by \eqref{sun33}.

\subsection{{Massive vector field equations}}

Although it is not the main topic of this paper, let us briefly comment on a generalization of the Maxwell field to the massive case. The vector Proca field $\ts{\EMA}$ satisfies the following field equations \cite{Proca:1936,Belinfante:1949,Rosen:1994,Seitz:1986sc}:
\begin{equation}\label{ProcaEq}
    \nabla_{\!n}\EMF^{an} + m^2 \EMA^a = 0 \;.
\end{equation}
As a direct consequence, the Lorentz condition \eqref{LorCond} must be satisfied. Employing \eqref{MaxwellZ}, \eqref{ProcaEq} for our ansatz gives
\begin{equation}\label{ProcaZvac}
    B^{am}\nabla_{\!m}\Bigl(\Box Z
    +2\beta \xi_k B^{kn}\nabla_{\!n}Z\Bigr) =
    m^2 B^{am}\nabla_{\!m} Z
    \;.
\end{equation}
Clearly, it is satisfied if
\begin{equation}\label{ProcaEqEigenfc}
    \Bigl[\Box +2\beta \xi_k B^{kn}\nabla_{\!n}\Bigr]Z = m^2\,Z
    \;.
\end{equation}
The sufficient conditions for the Proca equations are thus the Lorentz condition \eqref{LorCond} and the eigenfunction equation \eqref{ProcaEqEigenfc} of the operator \eqref{BoxModOp}.

In coordinates, the previous results \eqref{nablaBnablaZcoor} and \eqref{BoxModOpcoor} require
\begin{align}
    \frac1{A}\sum_\nu \frac{A_\nu}{U_\nu}\tilde{\sop}_\nu Z &= 0\;,\label{ProcaCoorLC}\\
    \sum_\nu \frac{1}{U_\nu}\tilde{\sop}_\nu Z &= m^2\, Z\;.\label{ProcaCoorPE}
\end{align}
The electromagnetic case is recovered upon switching off the mass, $m^2=0$.

\section{Structure of the equations}
\label{sc:StrcEqs}

In this section we are going to discuss the general structure of the obtained equations, the associated system of commuting operators, and the corresponding eigenvalue problem. We start with the following observation.

\subsection{{\boldmath$k\textendash\nu$} transform}

{Let $O_k$, $k=0,\dots, \dg-1$} be a set of $\dg$ `objects'. Define the following `polynomials':
\be
\tilde O_\nu=\sum_k (-x_\nu^2)^{\dg-1-k}O_k\,.
\ee
{$\tilde O_\nu$ are thus polynomials in variable $x_\nu^2$ with the same coefficients $O_k$.}
Applying the algebraic relation \eqref{Aid1i} we can write
\be
O_k=\sum_\nu \frac{A_\nu^{(k)}}{U_\nu}\tilde O_\nu\,.
\ee
Moreover, we can define the following `generating' polynomial $O$, depending on an auxiliary variable $\beta$:
\be
O\equiv \sum_k O_k \beta^{2k}=\sum_\nu \frac{A_\nu}{U_\nu}\tilde O_\nu\,.
\ee
We can think of the above relations as {$(k\textendash\nu)$-transform between $O_k$ and ${\tilde O_\nu}$} objects.
In particular, if $O_k$ are ordinary numbers, $\tilde C_\nu$ are normal polynomials. In what follows, however, we will use this transform also for the differential operators. In such a case, $\tilde O_\nu$ will typically be an operator in variable $x_\nu$ only.

\subsection{System of commuting operators}

The Lorenz condition \eqref{nablaBnablaZcoor} and the modified box operator \eqref{BoxModOpcoor} are constructed using the same operators $\tilde{\sop}_\nu$ \eqref{sun33}.
Introducing the Killing-vector operators $\lop_k$,
\be
  \lop_k = -i\frac{\pa}{\pa\psi_k}\;,\label{lopkdef}
\ee
together (by employing the $(k\textendash\nu)$-transform) with the associated operators  $\tilde{\lop}_\nu$ and~$\lop$
\begin{gather}
  \tilde{\lop}_\nu = \sum_k (-x_\nu^2)^{\dg{-}1{-}k}\lop_k\;,\quad
  \lop_k = \sum_\nu \frac{\A{k}_\nu}{U_\nu}\, \tilde{\lop}_\nu\;,\label{lopmukrel}\\
  \lop = \sum_k \lop_k \beta^{2k}=\sum_\nu\frac{A_\nu}{U_\nu}\tilde{\lop}_\nu\;,\label{lopdef}
\end{gather}
the operators $\tilde{\sop}_\nu$ take the following form:
\begin{equation}\label{SymTilOp}
    \tilde{\sop}_\nu
    =(1{+}\beta^2x_\nu^2)\frac{\pa}{\pa x_\nu}
      \Bigl[\frac{X_\nu}{1{+}\beta^2x_\nu^2}\frac{\pa}{\pa x_\nu}\Bigr]
    - \frac1{X_\nu} \tilde{\lop}_\nu^2
    +i\beta\frac{1{-}\beta^2x_\nu^2}{1{+}\beta^2x_\nu^2}\;
      \beta^{2(1-\dg)}\lop\;.
\end{equation}
%
In a similar manner, starting from $\tilde{\sop}_\nu$, we introduce $\sop_k$ and~$\sop$,
\begin{gather}
  \tilde{\sop}_\nu = \sum_k (-x_\nu^2)^{\dg{-}1{-}k}\sop_k\;,\quad
  \sop_k = \sum_\nu \frac{\A{k}_\nu}{U_\nu}\, \tilde{\sop}_\nu\;,\label{sopmukrel}\\
  \sop = \sum_k \sop_k \beta^{2k}=\sum_\nu\frac{A_\nu}{U_\nu}\tilde{\sop}_\nu\;.\label{sopdef}
\end{gather}
Using these definitions, we can present the operators {\eqref{nablaBnablaZcoor}} and {\eqref{BoxModOpcoor}} in the following form:
{\begin{align}
\nabla_{\!m}\bigl[B^{mn}\nabla_{\!n}\bigr]
    =\frac1{A} &\sum_\nu \frac{A_\nu}{U_\nu}\,\tilde\sop_\nu
    =\frac1{A}\sop
    \;,\label{nablaBnablaZstr}\\
\bigl[\Box +2\beta \xi_k B^{kn}\nabla_{\!n}\bigr]
    = &\sum_\nu \frac{1}{U_\nu}\,\tilde\sop_\nu
    = \sop_0
    \;.\label{BoxModOpstr}
\end{align}}

It is important to notice that unlike operators $\lop_k$ and $\tilde{\lop}_\mu$, operators  $\sop_k$ and $\tilde{\sop}_\mu$ are $\beta$-dependent. We do not write this dependence explicitly but we should remember it. In case of the Killing-vector operator $\lop$ the expansion in~$\beta$ gives directly $\lop_k$. Since $\sop_k$ depend on $\beta$, the same is not true for $\beta$-expansion of $\sop$, although the relation \eqref{sopdef} still holds true.

{Let us observe here that the operator $\nabla_{\!m}\bigl[B^{mn}\nabla_{\!n}\bigr]$ is symmetric for $\beta$ imaginary. This follows from the fact that $\beta$ enters the definition of $\ts{B}$ in a combination with antisymmetric $\ts{h}$, cf.~\eqref{Bdef}.
This might suggest we set
\begin{equation}\label{betamu}
\beta=-i\emp\;,
\end{equation}
assuming $\emp$ to be real; this notation has been used in \cite{FrolovKrtousKubiznak:2018a}. However, such a choice would possibly restrict the ability to describe a sufficient set of independent polarizations as can be seen from discussion in \cite{Frolov:2018ezx}. For this reason in what follows we continue working with a general complex~$\beta$.}

An important property of the operators $\sop_k$ and $\lop_k$ is that, for a fixed value of $\beta$, these operators mutually commute
\begin{equation}\label{soplopcom}
  [\sop_k,\sop_l]=0\;,\quad [\sop_k,\lop_l]=0\;,\quad [\lop_k,\lop_l]=0\;.
\end{equation}
Beware, however, that for different values of $\beta$, this is no longer true, and ${[\sop_k(\beta_1),\sop_l(\beta_2)]\neq0}$.

The commutation of the Killing-vector operators ${\lop_k}$ is obvious. The commutation of $\sop_k$ follows from the commutation of operators $\tilde{\sop}_\nu$. Each $\tilde{\sop}_\nu$ contains just one $x$-variable $x_\nu$, derivatives with respect to $x_\nu$, and derivatives with respect to all $\psi_k$. Therefore, for $\kappa\neq\lambda$ operators $\tilde{\sop}_\kappa$ and $\tilde{\sop}_\lambda$ trivially commute. For $\kappa=\lambda$ they commute only for the same value of $\beta$, when they are identical. For fixed $\beta$, we thus have $\tilde{\sop}_\kappa\tilde{\sop}_\lambda=\tilde{\sop}_\lambda\tilde{\sop}_\kappa$. Expanding the right operators using \eqref{sopmukrel} we get
\begin{equation}
  \sum_l\Bigl((-x_\lambda^2)^{\dg{-}1{-}l}\tilde{\sop}_\kappa\sop_l
    +\delta^\kappa_\lambda H_\kappa^l\sop_l\Bigr)
  =\sum_k\Bigl((-x_\kappa^2)^{\dg{-}1{-}k}\tilde{\sop}_\lambda\sop_k
    +\delta^\lambda_\kappa H_\lambda^k\sop_k\Bigr)\;,
\end{equation}
where $H_\nu^j=\tilde{\sop}_\nu (-x_\nu^2)^{\dg{-}1{-}j}$. Sums of terms with $H$'s on both sides are the same. Applying relation \eqref{sopmukrel} once more we get
\begin{equation}
  \sum_{k,l}(-x_\kappa^2)^{\dg{-}1{-}k}(-x_\lambda^2)^{\dg{-}1{-}l}\,\sop_k\sop_l
  =\sum_{k,l}(-x_\kappa^2)^{\dg{-}1{-}k}(-x_\kappa^2)^{\dg{-}1{-}l}\,\sop_l\sop_k\;.
\end{equation}
Since the matrix $(-x_\nu^2)^{\dg{-}1{-}j}$ (indexed by $\nu$ and $j$) is nonsingular, the commutativity ${[\sop_k,\sop_l]=0}$ follows.

\subsection{System of eigenfunctions}

For fixed $\beta$ we have commuting operators ${\sop_k}$ and ${\lop_k}$. We can thus introduce a system of common eigenfunctions ${\efc\equiv\efc(\beta;\scs_0,\dots,\scs_{\dg{-}1},L_0,\dots,L_{\dg{-}1})}$ labeled by eigenvalues ${\scs_k}$ and ${L_k}$,
\begin{equation}\label{eigenfc}
\begin{aligned}
   \sop_k\, \efc &= \scs_k \efc\;,\\
   \lop_k\, \efc &= L_k \efc\;.
\end{aligned}
\end{equation}
The eigenvalues ${L_k}$ are related to the explicit symmetries corresponding to coordinates ${\psi_k}$. For periodic angular coordinates, such operators would acquire discrete values. We refer to the appendix \ref{sc:KillingAngles} for the corresponding discussion.

On the other hand, at the moment we do not have a covariant form for the operators ${\sop_k}$ which would connect the eigenvalues ${\scs_k}$ with some physical quantities. We expect that such operators are related to the hidden symmetries encoded by Killing tensors. Unfortunately, some obvious guesses for ${\sop_l}$ as  for example $\covd\cdot\KT{l}\cdot\covd -2i\beta \ts{l}_{(l)} \cdot B \cdot \covd$ do not quite work. Although the operators ${\sop_k}$ depend on ${\beta}$, we understand eigenvalues ${\scs_k}$, as well as ${L_k}$, to be \mbox{${\beta}$-independent}.\footnote{%
This is just a convention how to parameterize eigenfunctions $\efc$ for various values of ${\beta}$. Such a parametrization is possible if we assume that for different values of ${\beta}$ the spectrum of operators remains the same. This assumption may be too strong for the detailed study of the spectrum following from e.g. regularity of the eigenfunctions. Since we do not do such a study here, we can ignore potential problems and assume that, at least in some range of ${\beta}$, the spectrum remains the same, and that we can use the same eigenvalues for different ${\beta}$ to parameterize these eigenfunctions.}

In what follows, we shall use the eigenfunctions ${\efc}$ to generate solutions of the Maxwell equations.

Let us finish this section by introducing some auxiliary notation. Starting with constants ${\scs_k}$ and ${L_k}$, and using the ${(k\textendash\nu)}$-transform as we did for operators, we define polynomials ${\tilde{\scs}_\nu\equiv\tilde{\scs}_\nu(x_\nu)}$ and ${\scs\equiv\scs(\beta)}$, and ${\tilde{L}_\nu\equiv\tilde{L}_\nu(x_\nu)}$ and ${L\equiv L(\beta)}$ as follows
\begin{gather}
  \tilde{\scs}_\nu = \sum_k (-x_\nu^2)^{\dg{-}1{-}k}\scs_k\;,\quad
  \scs_k = \sum_\nu \frac{\A{k}_\nu}{U_\nu}\, \tilde{\scs}_\nu\;,\label{scsmukrel}\\
  \scs = \sum_k \scs_k \beta^{2k}=\sum_\mu\frac{A_\nu}{U_\nu}\tilde{\scs}_\nu\;,\label{scsdef}
\end{gather}
\begin{gather}
  \tilde{L}_\nu = \sum_k (-x_\nu^2)^{\dg{-}1{-}k} L_k\;,\quad
  L_k = \sum_\nu \frac{\A{k}_\nu}{U_\nu}\, \tilde{L}_\nu\;,\label{lscmukrel}\\
  L = \sum_k L_k \beta^{2k}=\sum_\mu\frac{A_\nu}{U_\nu}\tilde{L}_\nu\;.\label{lscdef}
\end{gather}

\section{Solutions: magnetic polarizations}
\label{sc:MagPol}

\subsection{Parametrization using polarizations}

As it was shown above, the solution of the Maxwell equations can be generated through the ansatz \eqref{Aansatz} by function ${Z}$ satisfying the Lorenz condition and the condition \eqref{MaxwellZvac}, which, using \eqref{nablaBnablaZstr} and \eqref{BoxModOpstr}, are
\begin{equation}\label{Maxwelleqsop}
    \sop Z = 0 \;,\quad \sop_0\, Z = 0\;.
\end{equation}

The second condition can be easily satisfied by our eigenfunctions ${\efc}$ with ${\scs_0=0}$. Note that the trivial ${\scs_0}$ is no longer required for the massive vector field discussed in the previous section, where we effectively require ${\scs_0=m^2}$, cf.~\eqref{ProcaCoorPE}; see \cite{Frolov:2018ezx} for a discussion of  interesting consequences. The first condition, when applied to ${\efc}$, requires
\be
C(\beta)\equiv\sum_k \scs_k \beta^{2k}=0\,.
\ee
This could be trivially satisfied by setting all ${\scs_k=0}$, but it would reduce our system of eigenfunctions too much. Fortunately, we can utilize here the freedom in parameter ${\beta}$ by setting it to one of ${\dg-1}$ roots ${\beta_0,\dots,\beta_{\dg{-}2\!}}$ of the polynomial ${C(\beta)}$. We thus define ${\dg-1}$ ``magnetic polarizations'' ${\efc_\mg^P}$, ${P=0,\dots,\dg-2}$, each labeled by ${2\dg-1}$ constants,
\begin{equation}\label{Zmgsol}
    \efc_\mg^P(\scs_1,\dots,\scs_{\dg{-}1},L_0,\dots,L_{\dg{-}1})
      = \efc(\beta_P;0,\scs_1,\dots,\scs_{\dg{-}1},L_0,\dots,L_{\dg{-}1})\;.
\end{equation}

Setting ${\scs_0=0}$ implies that one of the roots, say ${\beta_0}$, of the polynomial ${C(\beta)}$ is zero, ${\beta_0=0}$, cf.\ definition \eqref{scsdef}. However, for ${\beta=0}$ our ansatz \eqref{Aansatz} gives a pure gauge field. We thus have only ${\dg-2}$ magnetic polarizations ${P=1,\dots,\dg-2}$ corresponding to nonvanishing roots. We will discuss the missing ``electric polarization'' below.

Let us stress that we use the names ``magnetic'', ``electric'' and ``polarization'' in a very vague and intuitive way. We are motivated partially by a notation of Lunin \cite{Lunin:2017}, although the solutions above do not correspond directly to those introduced by Lunin. No direct relation to the traditional concepts of polarization is indicated here. We just expect that for massless field in ${D=2\dg}$ dimensions one should have ${D-2}$ polarizations (or ${\dg-1}$ complex polarizations), each labeled by ${2\dg-1}$ constants. At this moment we found ${\dg-2}$ such complex polarizations.

\subsection{Alternative parametrization}

In the picture described above we use eigenvalues ${\scs_k}$, ${L_k}$ to parameterize solutions and for each choice of them we set ${\beta}$ to one of the roots ${\beta_P}$. It gives us a discrete choice of the polarization for given eigenvalues. Clearly, changing constants ${\scs_k}$, ${L_k}$ varies the roots ${\beta_P}$ and these roots can mix their values. The nature of independence of the polarizations is thus not completely clear.

One can therefore prefer a different parametrization of functions satisfying the Lorenz condition ${\sop Z=0}$. Instead of constants ${\scs_0,\dots,\scs_{\dg{-}1}}$, which define a root ${\beta_*}$, one can use the root ${\beta_*}$ and constants ${\scs_1,\dots,\scs_{\dg{-}1}}$ as independent and find a value of ${\scs_0}$ so that ${\scs(\beta_*)=0}$. Clearly, ${\scs_0}$ must be given by
\begin{equation}\label{scs0bardef}
    \bar{\scs}_0 = -\beta_*^2 \sum_{k=0}^{\dg-2}\scs_{k+1}\beta_*^{2k}\;.
\end{equation}
The Maxwell equations then require ${\bar\scs_0 = 0}$. It can be achieved by setting ${\beta_*=0}$, which leads to a pure gauge as before, or by imposing a linear constraint
\begin{equation}\label{mgaltMaxwell}
  \sum_{k=0}^{\dg-2}\scs_{k+1}\beta_*^{2k} = 0
\end{equation}
on the remaining  constants ${\scs_1,\dots,\scs_{\dg{-}1}}$. It can be solved, for example, by evaluating ${\scs_1}$ in terms of other constants,
\begin{equation}\label{scs1bardef}
    \bar{\scs}_1 = -\beta_*^2 \sum_{k=0}^{\dg-3}\scs_{k+2}\beta_*^{2k}\;.
\end{equation}

The ``magnetic'' solutions of the Maxwell equations can thus be generated through the ansatz \eqref{Aansatz} using functions
\begin{equation}\label{Zmgaltsol}
    \efc_{\mg}(\beta_*;\scs_2\dots,\scs_{\dg{-}1},L_0,\dots,L_{\dg{-}1})
      = \efc(\beta_*;0,\bar{\scs}_1,\scs_2,\dots,\scs_{\dg{-}1},L_0,\dots,L_{\dg{-}1})\;.
\end{equation}
In this parametrization we do not distinguish a discrete choice of the polarization, instead we have a direct control over the root ${\beta_*}$. However, changing ${\beta_*}$ and $\scs_2,\dots,\scs_{\dg{-}1},L_0,\dots,L_{\dg{-}1}$ freely should cover the same set of function as ${\efc_\mg^P}$ introduced above.

This parametrization corresponds more to Lunin's approach, as far as we are able to compare.

\subsection{Yet another parametrization}

Another way how to solve the Lorenz condition, i.e., to enforce that ${\beta_*}$ is a root of ${\scs(\beta)}$, is to require that the polynomial ${C}$ has the following form:
\begin{equation}\label{scsintermsofQ}
    C = (\beta^2-\beta_*^2)\, Q\;, \quad Q = \sum_{k=0}^{\dg-2} Q_k\, \beta^{2(\dg{-}2{-}k)}\;.
\end{equation}
It gives ${C_k}$ in terms of ${Q_0,\dots,Q_{\dg{-}2}}$ and ${\beta_*}$, namely for~${\scs_0}$,
\begin{equation}\label{mgalt2scs0}
    \scs_0 = -\beta_*^2\, Q_{\dg{-}2}\;.
\end{equation}
Setting thus ${Q_{\dg{-}2}=0}$ guarantees the Maxwell equations. Hence, the solution is generated by function~${\efc_{\mg'}}$
\begin{equation}\label{Zmgalt2sol}
  \efc_{\mg'}(\beta_*;Q_0\dots,Q_{\dg{-}3},L_0,\dots,L_{\dg{-}1})
      = \efc(\beta_*;0,\scs_1,\scs_2,\dots,\scs_{\dg{-}1},L_0,\dots,L_{\dg{-}1})\;,
\end{equation}
with ${\scs_k}$ evaluated from ${\beta_*}$ and ${Q_0,\dots,Q_{\dg{-}3}}$ using \eqref{scsintermsofQ}.

It will be useful in a discussion of the separation of variables to evaluate polynomials ${\tilde{\scs}_\nu}$ in this parametrization. Employing \eqref{scsintermsofQ} and \eqref{scsmukrel}, one easily gets
\begin{equation}\label{tildescsintermsofQ}
    \tilde{\scs}_\nu = 
    (1+\beta_*^2x_\nu^2)\, \bar{Q}_\nu\;,
\end{equation}
where we have introduced polynomials ${\bar{Q}_\nu\equiv \bar{Q}_\nu(x_\nu)}$
\begin{equation}\label{Qnudef}
    \bar{Q}_\nu = \sum_{k=0}^{\dg-2} Q_k (-x_\nu^2)^k\;,
\end{equation}
with the highest power missing when the Maxwell equations are imposed.

\subsection{D=4}

The last parametrization is suitable for a discussion of the four-dimensional spacetimes. Namely, for ${\dg=2}$, both polynomials ${Q}$ and ${\bar{Q}_\nu}$ reduce to a constant and the Maxwell equations require this constant to be zero. The solution is generated by the function ${\efc_{\mg'}(\beta_*,L_0,L_1)}$ parameterized just by the root ${\beta_*}$ and Killing-vector constants ${L_0,L_1}$. Clearly, ${\scs_k=0}$, as well as ${\tilde{\scs}_\nu=0}$, and ${\beta_*}$ is unconstrained.

\section{Solutions: electric polarization}
\label{sc:ElPol}

In the discussion of magnetic polarizations we have lost one solution, since ${\beta=0}$ leads to a pure gauge potential ${\ts{\EMA}=\covd Z}$. In this section we attempt to recover the missing polarization by investigating the behavior of our system of eigenfunctions \eqref{eigenfc} in the limit ${\beta\to0}$, see appendix~\ref{sc:betazero}. The obtained results inspire the following new ansatz for the vector potential:
\begin{equation}\label{AansatzEP}
    \EMA_a = h_{an} \nabla^n Z\;.
\end{equation}
The Lorenz condition and the Maxwell equations then read
\begin{gather}
   \nabla_{\!n}\EMA^n = (D-1) \xi^n\nabla_{\!n} Z = 0\;,\label{LCEP}\\
   \nabla_{\!n}\EMF^{an} = -h^{an}\nabla_{\!n}\Box Z +2 \xi^a \Box Z
     +(D-3) \nabla^a(\xi^n\nabla_{\!n}Z) = 0\;.\label{divFEP}
\end{gather}
Both these equations can be satisfied by requiring
\begin{gather}
    \Box Z = 0\;,\label{boxEP}\\
    \xi^n\nabla_{\!n} Z = 0\;.\label{statEP}
\end{gather}

The solutions of the wave operator have been studied before \cite{Frolov:2006pe,Sergyeyev:2007gf}. In appendix~\ref{sc:scWaveSep} we recall that they are given by the eigenfunctions ${\tilde{\efc}(K_0,\dots,K_{\dg{-}1},L_0,\dots,L_{\dg{-}1})}$ of the operators ${\kop_k}$ and ${\lop_k}$, labeled by their eigenvalues. The first condition \eqref{boxEP} sets the eigenvalue of the wave operator itself to zero, ${K_0=0}$. The second condition \eqref{statEP} requires ${L_0=0}$.

We can thus generate solutions to the Maxwell equations via ansatz \eqref{AansatzEP} using functions
\begin{equation}\label{EP}
    \efc_{\el}(K_1,\dots,K_{\dg{-}1},L_1,\dots,L_{\dg{-}1})
    = \tilde{\efc}(0,K_1,\dots,K_{\dg{-}1},0,L_1,\dots,L_{\dg{-}1})\;.
\end{equation}
We call these solutions the ``electric polarization''. This family of solutions is degenerate since it is parameterized just by ${2\dg-2}$ constants.

\section{Separation of variables}
\label{sc:SepVar}

Until now our discussion of the solutions of the Maxwell equations has been rather abstract, based  on the eigenfunctions of the operators ${\sop_k}$ and ${\lop_k}$. Here we demonstrate that these eigenfunctions can be found using the method of separation of variables. This reduces the problem to solving ordinary differential equations instead of having to deal with the complicated partial differential operators.

We proceed as follows. First, we show that the eigenvalue problem for the operators
${\sop_k}$ and ${\lop_k}$, and common eigenfunction $Z$, can be solved by employing the separation ansatz  \eqref{Zsep} below. Next, we discuss a refined method of separation of variables that is applicable to test fields  in the higher-dimensional Kerr--NUT--(A)dS spacetimes and show that both the Lorenz condition and the remaining Maxwell equations can be solved by this method. By comparing the obtained separated equations with the equations for the eigenfunction $Z$ we conclude that the separation constants are precisely the eigenvalues of the operators ${\sop_k}$ and ${\lop_k}$.

\subsection{Multiplicative separation ansatz}

A possibility to use the method of separation of variables is based on the fact that the operators ${\sop_k}$ and ${\lop_k}$ have a special form\footnote{%
The same structure has been already recognized in the discussion of the scalar field in \cite{Sergyeyev:2007gf} with operators ${\kop_k}$ and ${\tilde{\kop}_\nu}$, see\ appendix~\ref{sc:scWaveSep} for a short review. Naturally, we follow this case.}
\begin{equation}\label{sopsep}
    \sop_k = \sum_\nu \frac{\A{k}_\nu}{U_\nu} \tilde{\sop}_\nu\;,\quad
    {\lop_k}= \sum_\nu \frac{{\A{k}_\nu}}{U_\nu} \tilde{\lop}_\nu\,,
\end{equation}
where each ${\tilde{\sop}_\nu}$ and $\tilde{\lop}_\nu$ are operators in just one ${x}$-variable ${x_\nu}$ (and Killing variables $\psi_k$). We can take an advantage of this special coordinate dependence and of the additive structure by imposing the multiplicative separation ansatz for a function on which the operators act. Namely, we set
\begin{equation}\label{Zsep}
    Z = \Bigl(\prod_\nu R_\nu\Bigr) \; \exp\Bigl(i\sum_j L_j \psi_j\Bigr)\;,
\end{equation}
where functions $R_\nu$ are functions of just one variable, $R_\nu=R_\nu(x_\nu)$. Note that in terms of periodic angular coordinates $\ph_\nu$ \eqref{psiphirel} and constants $\mcs_\nu$ \eqref{mscdef}, the exponent reads
\begin{equation}\label{Lpsimph}
    \sum_k L_k \psi_k = \sum_\nu \mcs_\nu \ph_\nu\;.
\end{equation}

\subsection{Eigenvalue problem}
Let us consider the eigenvalue problem \eqref{eigenfc}, with the eigenfunction ansatz \eqref{Zsep}. The second set of equations, ${\lop_k}Z=L_k Z$, is automatically satisfied.  The first set of equations reads
\be\label{Ceigenfceq}
{\sop_k} Z={\scs_k}Z\,.
\ee
Writing the l.h.s. explicitly for the ansatz \eqref{Zsep}, we find
\begin{equation}\label{sopefcsep}
   {\sop_k} Z=Z\sum_\nu \frac{\A{k}_\nu}{U_\nu}\frac{1}{R_\nu}\biggl(
    (1{+}\beta^2x_\nu^2)\Bigl(\frac{X_\nu}{1{+}\beta^2x_\nu^2} R_\nu'\Bigr)'
    -\frac{{\tilde{L}}_\nu^2}{X_\nu}R_\nu
   +i\beta\frac{1{-}\beta^2x_\nu^2}{1{+}\beta^2x_\nu^2}\beta^{2(1-\dg)}L R_\nu\biggr)\,,
\ee
where the prime denotes the derivative with respect to ${x_\nu}$. Applying the $k\textendash\nu$ transformation to \eqref{Ceigenfceq}, the sum in \eqref{sopefcsep} disappears on the r.h.s. and polynomials $\tilde C_\nu$ defined in \eqref{scsmukrel} appear on l.h.s., yields  the following ordinary differential equations for functions $R_\nu$:
\begin{equation}\label{Rsepeqs}
   (1{+}\beta^2x_\nu^2)\Bigl(\frac{X_\nu}{1{+}\beta^2x_\nu^2} R_\nu'\Bigr)'
       -\frac{{\tilde{L}}_\nu^2}{X_\nu}R_\nu
   +i\beta\frac{1{-}\beta^2x_\nu^2}{1{+}\beta^2x_\nu^2}\beta^{2(1-\dg)}L R_\nu
    -\tilde{\scs}_\nu R_\nu =0\;.
\end{equation}

Functions ${R_\nu}$, each of one variable ${x_\nu}$, satisfying equations \eqref{Rsepeqs} thus give eigenfunctions ${\efc(\beta,\scs_0,\dots,L_0,\dots)}$ via multiplicative ansatz \eqref{Zsep}.

\subsection{Refined separation of variables}

We now want to demonstrate that the eigenvalues, which label our eigenfunctions, can be interpreted as separation constants. We start by describing the refined method of separation of variables that is applicable in our case.

An elementary formulation of the method of separation of variables states that if one has ${\dg}$ functions ${f_\nu}$, each of which depends on one variable only, ${f_\nu=f_\nu(x_\nu)}$, and if they add to zero, ${\sum_\nu f_\nu=0}$, then each $f_\nu$ has to be a constant and these constants have to sum to zero,
\begin{equation}\label{sepexample}
    f_\nu = q_\nu\;,\quad \sum_\nu q_\nu = 0\;.
\end{equation}
${q_\nu}$ are called separation constants and only ${\dg-1}$ of them are independent.

In the following we use a slightly different notion of separability. We formulate it as:\\
\textbf{Separation lemma.} \textit{Let ${f_\nu}$ are ${\dg}$ functions of one variable only, $f_\nu=f_\nu(x_\nu)$. If they composite to a zero according to
\begin{equation}\label{sepHD0}
    \sum_\nu \frac{1}{U_\nu} f_\nu = 0\;,
\end{equation}
then they have to be given by the same polynomial of degree ${\dg-2}$:
\begin{equation}\label{fsepQ}
    f_\nu = \sum_{k=0}^{\dg-2} Q_k (-x_\nu^2)^k\equiv \bar{Q}_\nu\;.
\end{equation}
We call the coefficients of these polynomials, ${Q_0,\dots,Q_{\dg{-}2}}$, the separation constants. There are ${\dg-1}$ of them, one less than the number of independent variables.}

The implication \eqref{fsepQ} to \eqref{sepHD0} follows directly from identity \eqref{Aid1i}. The proof of the opposite implication has been sketched in \cite{Krtous:2007}.

The lemma encodes the greatest freedom in functions ${f_\nu}$ which compose to zero through the sum of type \eqref{sepHD0}. It can be easily generalized to a non-trivial right hand side if one knows at least one particular solution ${f_\nu}$ for that right hand side. Namely, using again the identity \eqref{Aid1i}, we find:\\
\textbf{Generalized separation lemma.} \textit{Functions $f_\nu$ of one variable satisfying
\begin{equation}\label{sepHDconst}
    \sum_\nu \frac{1}{U_\nu} f_\nu = \scs_0\,,
\end{equation}
with $\scs_0=\text{const}$, must be given by a polynomial of degree ${\dg-1}$,
\begin{equation}\label{fsepC}
    f_\nu = \sum_{k=0}^{\dg-1} \scs_k (-x_\nu^2)^{\dg{-}1{-}k}\equiv\tilde{\scs}_\nu\;,
\end{equation}
where the constant ${\scs_0}$ specifies the highest order term.}

\subsection{Separation constants}

Using this insight, we can revisit the Lorenz condition ${\frac{1}{A}\sop Z =0}$, cf.~\eqref{nablaBnablaZstr}. Employing the separation ansatz \eqref{Zsep}, it yields
\begin{equation}\label{LCsep}
   \frac{1}{A}\sum_\nu \frac{A_\nu}{U_\nu}\frac{1}{R_\nu}\biggl(
    (1{+}\beta^2x_\nu^2)\Bigl(\frac{X_\nu}{1{+}\beta^2x_\nu^2} R_\nu'\Bigr)'
    -\frac{{\tilde{L}}_\nu^2}{X_\nu}R_\nu
   +i\beta\frac{1{-}\beta^2x_\nu^2}{1{+}\beta^2x_\nu^2}\beta^{2(1-\dg)}L R_\nu\biggr)
    = 0 \;.
\end{equation}
At first sight this equation does have the form \eqref{sepHD0} useful for the separation lemma since ${A_\nu}$ is a function of all variables $\{x_\kappa\}$ except ${x_\nu}$ and we need the exact opposite. Fortunately, the definition \eqref{Amudef} of ${A_\nu}$ shows that ${A_\nu/A}=(1+\beta^2x_\nu^2)^{-1}$ is function of just ${x_\nu}$. The Lorenz condition thus takes the form \eqref{sepHD0} where
\begin{equation}\label{LCsep}
   \sum_\nu \frac{1}{U_\nu}\frac{1}{R_\nu}\biggl(
    \Bigl(\frac{X_\nu}{1{+}\beta^2x_\nu^2} R_\nu'\Bigr)'
    -\frac{{\tilde{L}}_\nu^2}{(1{+}\beta^2x_\nu^2) X_\nu}R_\nu
   +i\beta\frac{1{-}\beta^2x_\nu^2}{(1{+}\beta^2x_\nu^2)^2}\beta^{2(1-\dg)}L R_\nu\biggr)
    = 0 \;,
\end{equation}
and the separation lemma gives
\begin{equation}\label{LCsepR}
   (1{+}\beta^2x_\nu^2)\Bigl(\frac{X_\nu}{1{+}\beta^2x_\nu^2} R_\nu'\Bigr)'
    -\frac{{\tilde{L}}_\nu^2}{X_\nu}R_\nu
   +i\beta\frac{1{-}\beta^2x_\nu^2}{1{+}\beta^2x_\nu^2}\beta^{2(1-\dg)}L R_\nu
    -(1{+}\beta^2x_\nu^2)\bar{Q}_\nu R_\nu = 0 \;.
\end{equation}

On the other hand, the remaining Maxwell equations reduce to $\sop_0 Z =0$, cf.\ \eqref{BoxModOpstr}. Slightly more generally, we {can consider the eigenfunction equation}
\begin{equation}\label{sop0sep}
    \sop_0\, Z = \scs_0 Z\;,
\end{equation}
setting $\scs_0=0$ later. Substituting the multiplicative separation ansatz \eqref{Zsep}, we obtain
\begin{equation}\label{sopefcsep0}
   \sum_\nu \frac{1}{U_\nu}\biggl(
    (1+\beta^2x_\nu^2)\Bigl(\frac{X_\nu}{1+\beta^2x_\nu^2} R_\nu'\Bigr)'
    -\frac{{\tilde{L}}_\nu^2}{X_\nu}R_\nu
   +i\beta\frac{1-\beta^2x_\nu^2}{1+\beta^2x_\nu^2}\beta^{2(1-\dg)}L R_\nu\biggr)
    =\scs_0 \;.
\end{equation}
The generalized separation lemma \eqref{sepHDconst} above then implies that the brackets must be equal to the same polynomial ${\tilde{\sop}_\nu}$ in the respective variable ${x_\nu}$,
\begin{equation}\label{sop0sepR}
   (1+\beta^2x_\nu^2)\Bigl(\frac{X_\nu}{1+\beta^2x_\nu^2} R_\nu'\Bigr)'
    -\frac{{\tilde{L}}_\nu^2}{X_\nu}R_\nu
   +i\beta\frac{1-\beta^2x_\nu^2}{1+\beta^2x_\nu^2}\beta^{2(1-\dg)}L R_\nu
    -\tilde{\scs}_\nu R_\nu = 0 \;.
\end{equation}
The eigenvalue $\scs_0$ determine the highest order of the polynomials $\tilde{\scs}_\nu$ and, as we said, the source-free Maxwell equations require $\scs_0=0$. We also realize that the separated equations \eqref{sop0sepR} are identical to the conditions \eqref{Rsepeqs} obtained from the eigenfunction equations for operators $\sop_k$. It means that the separability constants $\scs_k$ (coefficients of the polynomials $\tilde{\scs}_\mu$ from the generalized separation lemma) are exactly the eigenvalues of operators $\sop_k$.

Moreover, by comparing \eqref{LCsepR} and \eqref{sop0sepR}, we recover the relation \eqref{tildescsintermsofQ},
\begin{equation}\label{scsQrel}
    \tilde{\scs}_\nu = (1+\beta^2x_\nu^2)\bar{Q}_\nu\;,
\end{equation}
which we derived originally from a completely different perspective. However, the basis for this relation remains the same. It reflects the requirement that the Lorenz condition ${\sop Z =0}$ holds for given ${\beta}$.

\section{Aligned electromagnetic fields}\label{sc:aligned}

Let us  return to the electric polarization discussed in section \ref{sc:ElPol}. One can apply the multiplicative separation ansatz as in the previous section and recover the separable structure of the eigenfunctions ${\tilde{\efc}}$, see appendix~\ref{sc:scWaveSep}.

However, we will look at this case from a different perspective, restricting to the special case
\begin{equation}\label{Lk0}
  L_k=0\;,
\end{equation}
i.e., to the field independent of ${\psi_k}$.

In section \ref{sc:ElPol} we mentioned that the ansatz \eqref{AansatzEP} for the vector potential and the field equations \eqref{boxEP}, \eqref{statEP} can be motivated by the limiting procedure $\beta\to 0$ discussed in appendix \ref{sc:betazero}. It is then natural to assume that functions ${R_\nu}$ in the multiplicative separation ansatz \eqref{Zsep} also expand as
\begin{equation}\label{Rexp}
    R_\nu = 1 + \beta S_\nu + \mathcal{O}(\beta^2)\;,
\end{equation}
with functions ${S_\nu}$ depending just on one variable, ${S_\nu=S_\nu(x_\nu)}$.
The multiplicative separation ansatz thus reduces to
\begin{equation}\label{Zsepspecexp}
    Z = 1 + \beta \sum_\nu S_\nu + \mathcal{O}(\beta^2)\;.
\end{equation}

It motivates us to search for the electric polarization \eqref{AansatzEP},
\begin{equation}\label{AansatzEPS}
    \EMA_a = - h_{an}\nabla^n S\;,
\end{equation}
in the form of an additive separation ansatz
\begin{equation}\label{Zsepadd}
    S = \sum_\nu S_\nu\;.
\end{equation}
The vector potential yields
\begin{equation}\label{ASadd}
    \EMA_a = - h_{an} \sum_\nu S_\nu' \nabla^n x_\nu
    = \sum_\nu \frac{x_\nu X_\nu S_\nu'}{U_\nu}  \sum_k \A{k}_\nu d_a\psi_k\,.
\end{equation}

The Lorenz condition \eqref{statEP} is satisfied automatically since ${L_0=0}$. The Maxwell condition \eqref{boxEP} is the massless scalar wave equation ${\kop_0 S = 0}$, and upon inserting the additive separation ansatz, we get
\begin{equation}\label{boxScoor}
    \sum_\nu \frac{1}{U_\nu} \bigl(X_\nu S_\nu'\bigr)'  = 0\;.
\end{equation}
The separability lemma gives us that ${S_\nu}$ must satisfy the following differential equation:
\begin{equation}\label{Seqs}
    \bigl(X_\nu S_\nu'\bigr)' = \bar{Q}_\nu \;,
\end{equation}
where ${\bar{Q}_\nu}$ are polynomials \eqref{fsepQ} of degree ${\dg-2}$ in ${x_\nu^2}$. Eq.~\eqref{Seqs} can be integrated once, leading to
\begin{equation}\label{Seqsint}
    x_\nu X_\nu S_\nu' = q_\nu x_\nu + \tilde{P}_\nu  \;,
\end{equation}
where ${q_\nu}$ is an integration constant and ${\tilde{P}_\nu}$ is a polynomial of degree ${\dg-1}$ in ${x_\nu^2}$ without an absolute term, say
\begin{equation}\label{Ppol}
    \tilde{P}_\nu = \sum_{l=0}^{\dg-2} P_l (-x_\nu^2)^{\dg-1-l}\;.
\end{equation}
Substituting to the vector potential \eqref{ASadd}, and using \eqref{Aid1i}, we obtain
\begin{equation}\label{alignedfield}
    \ts{\EMA} 
     = \sum_\nu \frac{ q_\nu x_\nu }{U_\nu}  \sum_{k=0}^{\dg-1} \A{k}_\nu \grad\psi_k
     +  \sum_{k=0}^{\dg-2} P_k \,\grad \psi_k \,.
\end{equation}
The second term is a pure gauge can be ignored. The first term reproduces exactly the electromagnetic fields aligned with the principal tensor found in \cite{Krtous:2007} and discussed in~\cite{Kolar:2015cha}. In other words, we have just demonstrated that the aligned fields can be understood as a special case of the electric solutions for which the dependence on all Killing coordinates vanishes.

\section{Summary}
\label{sc:summary}


In this paper we have demonstrated the separability of the Maxwell equations in the background of the most general higher-dimensional spacetime admitting the principal tensor---the off-shell Kerr--NUT--(A)dS geometry. This goal was achieved by adopting a special ansatz \eqref{Aansatz} for the vector potential of the electromagnetic field. We demonstrated that this ansatz solves the Maxwell equations if the corresponding potential function $Z$ has the form \eqref{Zsep} provided mode functions $R_{\nu}$ are solutions of the second-order ODEs \eqref{Rsepeqs}. These equations contain the metric functions $X_{\nu}$. For a general off-shell metric, these are arbitrary functions of one variable $x_{\nu}$. For the on-shell metric these functions become polynomials, so that the coefficients that enter the equations are rational functions of the corresponding variables.

In order to adapt the obtained separated equations to the physical Kerr--NUT--(A)dS black hole spacetimes (in Lorentzian signature), one needs to {apply additionally} the Wick rotation to the radial coordinate and the mass parameter \cite{FrolovKrtousKubiznak:2017review}. Consequently, one of the separated equations will be in the radial sector, while the other  equations become the latitudinal angle equations. The requirement that the solutions of these angular equations are regular fixes the spectrum of some of the separation constants. We did not discuss these important details in the present paper, but we would like to emphasize that the proof of the completeness of the set of the solutions, obtained by in this paper  described method is an important open problem.

We also demonstrated that for the constructed potential function the electromagnetic field potential satisfies the Lorenz gauge condition.

Let us emphasize that the approach used in this paper is in its spirit  similar to the one proposed by Lunin \cite{Lunin:2017}. However, there are number of differences. First, we considered the off-shell Kerr--NUT--(A)dS metrics, and in this sense, we obtained a non-trivial and far-reaching generalization of Lunin's results. Second, contrary to Lunin's paper our construction is totally covariant and entirely based on the principal tensor. Third, the proof of the separability of the Maxwell equations proposed in our paper is carried out in an analytic form.

The key role in this proof is played by the rich geometrical structure generated from the principal tensor. Using this tensor we defined a covariant form of the polarization tensor which modifies the gradient of a generating scalar function in the ansatz for the vector potential. The rich symmetry structure has a consequence that both the Lorenz condition and the Maxwell equations can be written as a composition of operators separated in latitudinal variables. This allowed us to construct the system of commuting operators, which define a set of common eigenfunctions. In terms of these eigenfunctions we have been able to identify the separable solutions. These are labeled, in general, by the correct number of $D-1$ separation constants and we identified the correct number of polarizations.

Similar to the Klein--Gordon case, the obtained separated second-order ordinary differential equations for the potential function $Z$ can be identified with the eigenfunctions of a complete set of the first-order and second-order differential operators.
For the Klein--Gordon field, the covariant form of these operators is well known---they are constructed from the Killing vectors and Killing tensors present in the Killing tower. A covariant form for the operators acting on $Z$ is currently unknown and finding it poses an interesting problem for future studies.

\section*{Acknowledgments}
\label{sc:acknowledgements}

V.F.\ thanks the Natural Sciences and Engineering Research Council of Canada (NSERC) and the Killam Trust for their financial support, and thanks Charles University for hospitality.
P.K.\ was supported by Czech Science Foundation Grant 17-01625S.
D.K.\ acknowledges the Perimeter Institute for Theoretical Physics and the NSERC for their support. Research at Perimeter Institute is supported by the Government of Canada through the Department of Innovation, Science and Economic Development Canada and by the Province of Ontario through the Ontario Ministry of Research, Innovation and Science.



\appendix

\section{Angular coordinates}\label{sc:KillingAngles}

\subsection{Periodic angular coordinates}
In this appendix, let us return to the metric \eqref{metric} and discuss the meaning of the Killing coordinates ${\psi_k}$. Such coordinates correspond to explicit symmetries and represent time and longitudinal angles. However, this relation is not direct. Even for vanishing NUT parameters these coordinates cannot be directly identified with the standard periodic angular coordinates around axes of rotation but instead are their linear combination \cite{Chong:2004hw,Gibbons:2004js,FrolovKrtousKubiznak:2017review}. With non-vanishing NUTs, the situation is even more complicated because it is not clear, what are the ``correct'' periodic angular coordinates \cite{FrolovKrtousKubiznak:2017review,Kolar:2017vjl,Krtous:2015zco,KolarKrtous:2018}.

The reason is that the metric \eqref{metric} itself does not specify a global geometry. It has to be accompanied by a specification of what the ranges of coordinates are and which coordinates are periodic. In some cases (as vanishing NUTs, i.e., the Myers--Perry geometry) there is a natural choice of such angular coordinates which guarantees the regularity of axes of rotation.
With non-vanishing NUTs or even for the off-shell geometries described by \eqref{metric}, the axes cannot be, in general, made regular. Physically it means that there are some linear sources along the axes and such sources cannot be eliminated.\footnote{%
Even in the regular case of the Myers--Perry black hole one can superimpose rotating strings along the axes which is effectively done exactly be changing ranges of angular coordinates and rewinding angular and time coordinates among themselves. This causes irregularities on the axes. The regularity of Myers--Perry solution means that such irregularities can be eliminated by a proper choice of time and angular coordinates.}
The specification of these sources is hidden exactly in an identification of the periodic longitudinal coordinates.

In any case, the coordinates ${\psi_k}$ are not typically the periodic coordinates. ${\psi_0}$ is a time coordinate, but the periodic angular coordinates ${\ph_\nu}$ are given by a liner combination of ${\psi}$'s. In even ${2\dg}$ dimensions, which we mainly considered in the main text, it is useful to write such transformation in the form
\begin{equation}\label{psiphirel}
    \ph_\nu = \sum_{k=0}^{\dg-1}\frac{\Ac{k}_\nu}{\Uc_\nu}\psi_k\;,\quad
    \psi_k = \sum_\nu (-\xc_\nu^2)^{\dg{-}1{-}k}\ph_\nu\;,
\end{equation}
where ${\xc_\nu}$ are constants and all other quantitative as ${\Ac{k}_\nu}$, ${\Uc_\nu}$ are build from ${\xc_\nu}$ in the same way as ${\A{k}_\nu}$, ${U_\nu}$ from ${x_\nu}$. A specification of the constants ${\xc_\nu}$ thus defines the correct periodic coordinates ${\ph_\nu}$. It has to be accompanied by setting correct ranges of this periodicity. All these choices identify what singular sources are present on the axes.

For vanishing NUT parameters, the relation to the Myers--Perry metric includes setting ${\xc_\nu}$ to values of rotational parameters ${a_\nu}$, see \cite{FrolovKrtousKubiznak:2017review,Kolar:2017vjl,Krtous:2015zco}.

We will not discuss these issues in more detail. The only thing we need is that the periodic angular coordinates ${\ph_k}$ are related to ${\psi_k}$ by transformation \eqref{psiphirel}. Thus, when studying the spectra of operators related to Killing vectors, we can expect that operators ${\frac{\pa}{\pa\ph_\mu}}$ have a simple discrete spectrum and spectra of ${\frac{\pa}{\pa\psi_k}}$ must be derived from them using \eqref{psiphirel}. For that it is useful to write down relations of the coordinate Killing vectors:
\begin{equation}\label{psiphivecrel}
    \cv{\psi_k} = \sum_\nu\frac{\Ac{k}_\nu}{\Uc_\nu}\cv{\ph_\nu}\;,\quad
    \cv{\ph_\nu} = \sum_{k=0}^{\dg-1} (-\xc_\nu)^{\dg{-}1{-}k}\cv{\psi_k}\;,
\end{equation}
where we used the important identity \eqref{Aid1i}.

\subsection{Operators ${\mop_\nu}$ and ${\lop_k}$}

Operators ${\lop_k}$, ${\tilde{\lop}_\nu}$, and ${\lop}$ can be also expressed in terms of period coordinates ${\ph_\nu}$ introduced in \eqref{psiphirel}. If we define
\begin{equation}\label{mopdef}
  \mop_\nu = -i\frac{\pa}{\pa\ph_\nu}
  = \sum_k (-\xc_\nu^2)^{\dg{-}1{-}k}\lop_k\;,
\end{equation}
${\lop}$'s operators are
\begin{gather}
  \lop_k = \sum_\nu \frac{\Ac{k}_\nu}{\Uc_\nu}\, \mop_\nu\;,\label{lopkmoprel}\\
  \tilde{\lop}_\nu = \sum_\nu \frac{1}{\Uc_\nu}
    \prod_{\substack{\kappa\\\kappa\neq\nu}}(\xc_\kappa^2-x_\mu^2)\,\mop_\nu\;,\label{loptilmoprel}\\
  \lop = \sum_\nu\frac{\mathring{A}_\nu}{\Uc_\nu}\mop_\nu\;.\label{lopmoprel}
\end{gather}

Operators ${\mop_\nu}$ commute with all operators ${\sop_k}$ and ${\lop_k}$, since they are related to ${\lop_k}$ just by a linear transformation \eqref{mopdef} with constant coefficients. We can thus introduce eigenvalues of operators ${\mop_\nu}$,
\begin{equation}\label{mscdef}
  \mop_\nu\, \efc = \mcs_\nu \efc\;,
\end{equation}
which we expect to have a simple discrete spectrum.
The eigenvalues ${L_k}$ and polynomials ${\tilde{L}_\nu}$ and ${L}$ are then related as
\begin{gather}
  L_k = \sum_\nu \frac{\Ac{k}_\nu}{\Uc_\nu}\, \mcs_\nu\;,\label{lsckmscrel}\\
  \tilde{L}_\nu = \sum_\nu \frac{1}{\Uc_\nu}
    \prod_{\substack{\kappa\\\kappa\neq\nu}}(\xc_\kappa^2-x_\mu^2)\,\mcs_\nu\;,\label{lsctilmscrel}\\
  L = \sum_\nu\frac{\mathring{A}_\nu}{\Uc_\nu}\mcs_\nu\;.\label{lscmscrel}
\end{gather}
In the main text, we continue to use eigenvalues ${L_k}$ as basic ones. Transformation \eqref{lsckmscrel} can be always carried out at the end.

\section{Separability of the scalar wave equation}\label{sc:scWaveSep}

On several places in the main text we refer to the separability of the scalar wave equation in the off-shell Kerr--NUT--(A)dS spacetime. For convenience of the reader, in this appendix we give a short overview of this result. The separability has been first demonstrated in \cite{Frolov:2006pe} and later elaborated on in \cite{Sergyeyev:2007gf} where one can find also the results for odd dimensions. Here we restrict to even dimensions.

Using the Killing tensors $\KT{j}$ one can construct the tower of symmetric second order operators
\begin{equation}\label{kopdef}
    \kop_j = \nabla_{\!a}\bigl[k_j^{ab}\nabla_{\!b}\bigr]\;.
\end{equation}
In the canonical coordinates these operators read \cite{Sergyeyev:2007gf}
\begin{equation}\label{kopjcoor}
    \kop_j Z =\sum_\nu \frac{\A{j}_\nu}{U_\nu} \biggl[
    \frac{\pa}{\pa x_\nu}\Bigl[X_\nu\frac{\pa}{\pa x_\nu}\Bigr]
    + \frac1{X_\nu} \Bigl[{\sum_i (-x_\nu^2)^{\dg{-}1{-}i}\frac{\pa}{\pa\psi_i}}\Bigr]^2
    \biggr] Z\;.
\end{equation}
Similarly, the generating Killing tensor $\ts{k}$ defined in \eqref{KTfcexp}, defines the operator
\begin{equation}\label{kopcoor}
    \kop Z\equiv \nabla_{\!m}\bigl(k^{mn}\nabla_{\!n} Z\bigr)
    =\sum_\nu \frac{A_\nu}{U_\nu} \biggl[
    \frac{\pa}{\pa x_\nu}\Bigl[X_\nu\frac{\pa}{\pa x_\nu}\Bigr]
    + \frac1{X_\nu} \Bigl[\sum_j (-x_\nu^2)^{\dg{-}1{-}j}\frac{\pa}{\pa\psi_j}\Bigr]^2
    \biggr] Z\;.
\end{equation}

By a similar argument as for the operators $\sop_j$ in section \ref{sc:StrcEqs}, one can show that these operators, together with the operators $\lop_j$, mutually commute,
\begin{equation}\label{koplopcom}
  [\kop_k,\kop_l]=0\;,\quad [\kop_k,\lop_l]=0\;,\quad [\lop_k,\lop_l]=0\;.
\end{equation}
Therefore, they have common eigenfunctions $\bar\efc$ labeled by eigenvalues $K_j$ and $L_j$,
\begin{equation}\label{eigenfcKL}
\begin{aligned}
   \kop_j\, \bar\efc &= K_k \bar\efc\;,\\
   \lop_j\, \bar\efc &= L_k \bar\efc\;.
\end{aligned}
\end{equation}

Let us concentrate on the eigenfunction equation of the zeroth operator $\kop_0\equiv\Box$,
\begin{equation}\label{boxeveq}
  \kop_0 Z = K_0 Z\,.
\end{equation}
Substituting the multiplicative separation ansatz \eqref{Zsep}, it boils to
\begin{equation}\label{boxsepeq}
   \sum_\nu \frac{1}{U_\nu}\biggl(\frac{1}{R_\nu}
    \Bigl(X_\nu R_\nu'\Bigr)'
    -\frac{{\tilde{L}}_\nu^2}{X_\nu}\biggr)
    =K_0 \;.
\end{equation}
The generalized separation lemma \eqref{sepHDconst} then implies that the brackets must be equal to the same polynomial ${\tilde{K}_\nu}$ in the respective variable ${x_\nu}$ with $K_0$ governing the highest order term, i.e.,
\begin{equation}\label{boxsepeqR}
    \Bigl(X_\nu R_\nu'\Bigr)'
    -\frac{{\tilde{L}}_\nu^2}{X_\nu}R_\nu - \tilde{K}_\nu R_\nu
    =0 \;,
\end{equation}
where
\begin{equation}\label{fsepC}
    \tilde{K}_\nu \equiv \sum_{k=0}^{\dg-1} K_k (-x_\nu^2)^{\dg{-}1{-}k}\;.
\end{equation}

Plugging this most general solution of \eqref{boxeveq} to the operators ${\kop_j}$ we find that such ${Z}$ is the eigenfunction ${\bar\efc}$ with eigenvalues ${K_j}$. Using the system of eigenfunctions of the operators ${\kop_j}$ is thus equivalent to solving equation \eqref{boxeveq} by the separation of variables and the eigenvalues correspond to the separation constants.

Note also that by plugging $\bar\efc$ into operator \eqref{kopcoor}, we get
\begin{equation}\label{kopZ}
    \kop \bar\efc = K \bar\efc \;,
\end{equation}
with
\begin{equation}\label{Kgenfcdef}
    K = \sum_j K_j \beta^{2j}=\sum_\mu\frac{A_\nu}{U_\nu}\tilde{K}_\nu\;.
\end{equation}

\section{Limiting procedure $\beta\to 0$}\label{sc:betazero}

In the discussion of magnetic polarizations we have lost one of the solutions, since ${\beta=0}$ leads to a pure gauge potential ${\ts{\EMA}=\covd Z}$. In the hope to recover the missing polarization, let us investigate the behavior of our system of eigenfunctions \eqref{eigenfc} for ${\beta\to0}$.
As we shall see, this will naturally lead to the definition of the electric polarization \eqref{AansatzEP} in the main text. We shall also use this to recover the special solutions of Maxwell equations known as the aligned fields, see Sec.~\ref{sc:aligned}.

\subsection{Behavior of eigenfunctions for ${\beta\to0}$}

Observing the operators ${\tilde{\sop}_\nu}$ given by \eqref{SymTilOp}, we see a potential problem with the last term which is proportional to ${\beta^{2(1-\dg)}}$. To investigate its behavior, we expand also the fractional factor in ${\beta}$,
\begin{equation}\label{sopbeta0exp1}
  \frac{1-\beta^2x_\nu^2}{1+\beta^2x_\nu^2}\,\beta^{2(1-\dg)}
    = -1 +2\frac{1}{1+\beta^2x_\nu^2}\,\beta^{2(1-\dg)}
    =-1+2\sum_{j=0}^\infty \beta^{2(j{+}1{-}\dg)}(-x_\nu^2)^j\;.
\end{equation}
We see that the sum contains plenty of terms with negative powers of ${\beta}$. However, operators ${\sop_k}$ are given as a sum \eqref{sopmukrel} of operators ${\tilde{\sop}_\nu}$. Keeping just terms with non-positive powers of ${\beta}$ and changing ${j\to\dg-1-j}$, the contributions to ${\sop_k}$ are
\begin{equation}\label{sopbeta0exp2}
  \sum_\nu\frac{\A{k}_\nu}{U_\nu}\frac{1-\beta^2x_\nu^2}{1+\beta^2x_\nu^2}\,\beta^{2(1-\dg)}
    =-\sum_\nu\frac{\A{k}_\nu}{U_\nu}
    +2\sum_{j=0}^{\dg-1} \beta^{-2j} \sum_\nu\frac{\A{k}_\nu}{U_\nu} (-x_\nu^2)^{\dg{-}1{-}j} + \mathcal{O}(\beta^2)\;.
\end{equation}
Sums over ${\nu}$ are easily evaluated using the identity \eqref{Aid1i}, giving
\begin{equation}\label{sopbeta0exp3}
  \sum_\nu\frac{\A{k}_\nu}{U_\nu}\frac{1-\beta^2x_\nu^2}{1+\beta^2x_\nu^2}\,\beta^{2(1-\dg)}
    =-\delta_k^{\dg-1}\beta^{-2(\dg-1)} + 2\beta^{-2k} + \mathcal{O}(\beta^2)\;.
\end{equation}
Substituting it to \eqref{SymTilOp}, we get
\begin{equation}\label{sopbeta0exp}
    \sop_k = \kop_k + 2i\beta \sum_{l=0}^k \beta^{-2(k-l)} \lop_l + \mathcal{O}(\beta^2)
\end{equation}
for ${k=0,\dots,\dg-2}$ and
\begin{equation}\label{soptopbeta0exp}
    \sop_{\dg{-}1} = \kop_{\dg{-}1} + i\beta \sum_{l=0}^{\dg{-}1} \beta^{-2(\dg{-}1{-}l)} \lop_l+ \mathcal{O}(\beta^2)
\end{equation}
for ${k={\dg-1}}$. Here we used operators ${\kop_k}$ given in \eqref{kopjcoor}.

We see that the operators ${\sop_k}$ are mostly divergent for ${\beta\to0}$. Therefore, one cannot expect the eigenfunctions ${\efc}$ to behave reasonably in this limit. However, one could avoid this problem for a subclass of eigenfunctions, namely for those with ${L_k=0}$, i.e., for those independent of ${\psi_k}$. Operators ${\sop_k}$ acting on such eigenfunctions reduce just to ${\kop_k}$ (with the last term vanishing).

We can thus hope that by expanding the eigenfunctions ${\efc}$ with vanishing ${L_k}$ we could find a subfamily of solutions of the Maxwell equations corresponding to the missing polarization.

\subsection{Behavior of ${Z}$ for ${\beta\to0}$}

On other side, we are not obliged to use eigenfunctions~${\efc}$. We just want to solve \eqref{Maxwelleqsop}: ${\sop Z=0}$, and ${\sop_0 Z =0}$. Fortunately, ${\sop_0}$ is regular for small ${\beta}$,
\begin{equation}\label{sop0beta0exp}
    \sop_0 = \kop_0 + 2i\beta \lop_0 + \mathcal{O}(\beta^2)\;,
\end{equation}
as well as operator ${\sop}$,
\begin{equation}\label{sop0beta0exp}
    \sop = \sum_k \beta^{2k} \sop_k = \kop_0 + (2\dg{-}1)\,i\beta\, \lop_0 + \mathcal{O}(\beta^2)\;.
\end{equation}
Unfortunately, they differ in the first order of ${\beta}$ by a term proportional to ${\lop_0}$. Therefore, if we want satisfy both conditions \eqref{Maxwelleqsop} up to the first order, the function ${Z}$ must satisfy
\begin{gather}
    \kop_0 \, Z =0\;,\label{kop0cond}\\
    \lop_0 \, Z = 0\;.\label{lop0cond}
\end{gather}
The first condition is actually the scalar wave equation for ${Z}$. The second condition is milder than setting ${L_k=0}$ for all ${k}$ but it is still a non-trivial condition. The coordinate ${\psi_0}$ represents time, so we are obtaining the condition of stationarity.

\subsection{Behavior of field equations for ${\beta\to0}$}

Let us now look at the first-order expansion of the field equations. Assuming
\begin{equation}\label{Zexp}
    Z = Z_0 + \beta Z_1 + \mathcal{O}(\beta^2)\;,
\end{equation}
and ${\ts{B} = \ts{g} + \beta \ts{h} + \mathcal{O}(\beta^2)}$ following from \eqref{Bdef}, the ansatz \eqref{Aansatz} for the vector potential reads
\begin{equation}\label{Aexp}
    \ts{\EMA} =\covd Z_0 + \beta\bigl( \ts{h}\cdot\covd Z_0 + \covd Z_1 \bigr) + \mathcal{O}(\beta^2)\;.
\end{equation}
The leading term is a pure gauge, but the first order term is not. The nontrivial contribution comes from the zeroth-order function ${Z_0}$.

The Lorenz condition reads
\begin{equation}\label{LCexp}
   \covd\cdot\ts{\EMA} = \Box Z_0 + \beta \Bigl( (D-1)\ts{\xi}\cdot\covd Z_0 + \Box Z_1 \Bigr) + \mathcal{O}(\beta^2)\;,
\end{equation}
where we have used \eqref{PrimaryVec} and the antisymmetry of the principal tensor ${\ts{h}}$. The Maxwell tensor ${\ts{\EMF}}$ is sensitive only to the gauge non-trivial part of the vector potential and therefore it is of the first-order,
\begin{equation}\label{Fexp}
   \ts{F}=\covd\wedge\ts{A} = \beta\, \covd\wedge\bigl(\ts{h}\cdot\covd Z_0\bigr)+ \mathcal{O}(\beta^2)\;.
\end{equation}

Calculating the Maxwell equations is more involved, but it essentially follows the same steps as in the case of arbitrary ${\beta}$ discussed in section \ref{sc:FieldEqs}. Alternatively, one can just expand \eqref{MaxwellZ}. It yields
\begin{equation}\label{divFexp}
    \covd\cdot\ts{\EMF} = \beta\Bigl( \ts{h}\cdot\covd\Box Z_0
      - 2\ts{\xi} \Box Z_0
      - (D-3) \covd(\ts{\xi}\cdot\covd Z_0)\Bigr)+\mathcal{O}(\beta^2)\;.
\end{equation}
We see that it can be set equal to zero up to the first-order in ${\beta}$ provided that
\begin{gather}
    \Box Z_0\equiv \kop_0 Z_0 = 0\;, \label{boxZ0}\\
    \ts{\xi}\cdot\covd Z_0 \equiv i \lop_0 Z_0 = 0\;.
\end{gather}
We recovered that the function ${Z_0}$ must satisfy the scalar wave equation and the stationarity condition, the results \eqref{kop0cond} and \eqref{lop0cond} above.

Let us note that the same result can be obtained if one assumed function ${Z}$ in the form
\begin{equation}\label{Zexpalt}
    Z = 1 + \beta Z_0 + \beta^2 Z_1 + \mathcal{O}(\beta^2)\;.
\end{equation}
The expansions of the Lorenz condition and of the Maxwell equations look exactly the same as in \eqref{LCexp} and \eqref{divFexp}, respectively, just with higher power of ${\beta}$.

\section{Technicalities \& Proofs}
\label{sc:Proofs}

In this appendix we gather some important technical results and present proofs that are referred to in the main text.

First we list some important identities for the symmetric polynomials:
\begin{gather}
\label{Aid1i}
    \sum_\mu\A{k}_\mu \frac{(-x_\mu^2)^{\dg{-}1{-}l}}{U_\mu}=\delta^k_l\;,\\
\label{Aid7}
    \sum_{\mu} \frac{\A{k}_\mu}{x_\mu^2 U_\mu} = \frac{\A{k}}{\A{\dg}}\;,\\
\label{SAisA}
    \A{\dg{-}1} = \sum_\mu\A{\dg{-}1}_\mu\;.
\end{gather}

\paragraph{Proof of \eqref{nablaBnablaZ}}\mbox{}\\[-2ex]

Employing expression \eqref{Bder} for the derivative of $\ts{B}$, symmetric part \eqref{Bevenodd} of $\ts{B}$, relation \eqref{Bkrel}, and definition \eqref{ldef}, we can write
\begin{equation}\label{proofnablaBnablaZ1}
\begin{split}
  &\nabla_{\!m}\bigl(B^{mn}\nabla_{\!n}Z\bigr)
  = B^{mn}\nabla_{\!(m}\nabla_{\!n)}Z + (\nabla_{\!m}B^{mn})\nabla_{\!n}Z\\
  &\;\; = \frac1A\,k^{mn}\nabla_{\!m}\nabla_{\!n}Z
    + \frac{\beta}{A}\bigl(k^a{}_a\, \xi_m B^{mn} - \xi_m k^{mn}\bigr)\nabla_{\!n} Z\\
  &\;\; = \frac1A\,\nabla_{\!m}\bigl(k^{mn}\nabla_{\!n}Z\bigr)
    - \frac1A\,(\nabla_{\!m}k^{mn})\nabla_{\!n}Z
    +\frac{\beta}{A}\Bigl(\frac{k^a{}_a}{A} l^n
       +\beta \frac{k^a{}_a}{A} l^m h_m{}^n-l^n\Bigr)\nabla_{\!n}Z\;.
\end{split}
\end{equation}
Substituting \eqref{kdiv}, and using \eqref{dA} we obtain
\begin{equation}\label{proofnablaBnablaZ2}
\begin{split}
  &\nabla_{\!m}\bigl(B^{mn}\nabla_{\!n}Z\bigr)
  = \frac1A\,\nabla_{\!m}\bigl(k^{mn}\nabla_{\!n}Z\bigr)\\
  &\mspace{32mu}-\frac1{A^2}(\nabla_{\!m}A)
      \bigl(k^{mn}-\frac12k g^{mn}\bigr)\,\nabla_{\!n}Z
      +\frac1{A}\,\Bigl(-\frac12\frac{k^a{}_a}{A}(\nabla^{n}A)
      +\beta \bigl(\frac{k^a{}_a}{A} -1\bigr)l^n\Bigr)\,\nabla_{\!n}Z\\
  &\;\; = \frac1A\,\nabla_{\!m}\bigl(k^{mn}\nabla_{\!n}Z\bigr)
      +\frac1{A}\,\Bigl(-\frac1{A}(\nabla_{\!m}A)k^{mn}
      +\beta \bigl(\frac{k^a{}_a}{A} -1\bigr)l^n\Bigr)\,\nabla_{\!n}Z\;,\\
  &\;\; = \nabla_{\!m}\bigl(\frac1A\,k^{mn}\nabla_{\!n}Z\bigr)
        +\frac{\beta}{A} \bigl(\frac{k^a{}_a}{A} -1\bigr)l^n\,\nabla_{\!n}Z\;,
\end{split}
\end{equation}
which is what we wanted to show.

\bigskip
\paragraph{Proof of \eqref{ka1id}}\mbox{}\\[-2ex]

Using identities \eqref{Aid1i}, \eqref{Aid7}, and \eqref{SAisA} for $k,l=\dg-1$, one has
\begin{equation}\label{Aid7klN-1}
     \sum_{\mu} \frac{\A{\dg{-}1}_\mu}{U_\mu} \Bigl(\frac{2}{x_\mu^2}-1\Bigr)
     =2\frac{\A{\dg{-}1}}{\A{\dg}} - 1
     =\frac{\sum_\mu2\A{\dg{-}1}_\mu}{\A{\dg}} - 1\;.
\end{equation}
Under substitution $x_\nu^2\to 1+\beta^2x_\nu^2$ functions $U_\mu$, $\A{\dg{-}1}_\mu$, and $\A{\dg}$ behave as $U_\mu\to\beta^{2(\dg{-}1)}U_\mu$, $\A{\dg{-}1}_\mu\to A_\mu$, and $\A{\dg}\to A$. It gives us the relation
\begin{equation}\label{ka1idproof1}
     \beta^{2{-}2\dg}\sum_{\mu} \frac{A_\mu}{U_\mu} \Bigl(\frac{2}{1+\beta^2x_\mu^2}-1\Bigr)
     =\frac1{A} \sum_\mu 2A_\mu - 1\;.
\end{equation}
On the right-hand side we identify expression \eqref{trkdef} for the trace $k^a{}_a$ of the generating Killing tensor and we obtain
\begin{equation}\label{ka1idproof2}
    \frac{k^a{}_a}{A}-1 = \beta^{2{-}2\dg}\sum_{\mu} \frac{A_\mu}{U_\mu} \frac{1-\beta^2x_\mu^2}{1+\beta^2x_\mu^2}\;.
\end{equation}

\bigskip
\paragraph{Proof of \eqref{1overAid}}\mbox{}\\[-2ex]

We have
\begin{equation}\label{1overAid}
    \frac{\beta^{2(\dg{-}1)}}{A} = \sum_\nu\frac1{U_\nu}\frac1{1+\beta^2x_\nu^2}\;.
\end{equation}
Indeed, it is just \eqref{Aid7} with $k=0$, in which we substitute $x_\nu^2\to 1+\beta^2x_\nu^2$. Subtracting zero $0=\sum_\nu\frac1{U_\nu}$ (cf.~\eqref{Aid1i} with $k=0$ and $l=\dg-1$), we get
\begin{equation}\label{1overAid2}
    \frac{2}{A} = \beta^{2(1{-}\dg)}\sum_\nu\frac1{U_\nu}\frac{1-\beta^2x_\nu^2}{1+\beta^2x_\nu^2}\;.
\end{equation}

\bigskip
\paragraph{Proof of \eqref{boxA}}\mbox{}\\[-2ex]

We want to prove
\begin{equation}\label{boxA}
 \begin{split}
    &- \nabla_{\!m}\nabla^m (B^{an}\nabla_{\!n}Z) + R^a{}_m B^{mn}\nabla_{\!n}Z\\
    &\qquad\quad=-B^{an}\nabla_{\!n}\Box Z
    +2\beta B^{ak}\xi_k \nabla_{\!m}(B^{mn}\nabla_{\!n}Z)
    -2\beta B^{am}\nabla_{\!m}\bigl(\xi_k B^{kn}\nabla_{\!n}Z\bigr)
    \;.
 \end{split}
\end{equation}
We start by pulling $B^{an}$ from the covariant derivatives,
\begin{equation}\label{boxA1}
 \begin{split}
    - \nabla_{\!m}&\nabla^m (B^{an}\nabla_{\!n}Z) + R^a{}_m B^{mn}\nabla_{\!n}Z\\
    =&-B^{an}\nabla_{\!m}\nabla_{\!n}\nabla^m Z
    -2(\nabla^{m}B^{an})\nabla_{\!m}\nabla_{\!n}Z
    - (\nabla_{\!m}\nabla^{m}B^{an})\nabla_{\!n} Z
    + R^a{}_m B^{mn}\nabla_{\!n}Z\\
    =&-B^{an}\nabla_{\!n}\Box Z
    -2\beta\bigl(B^{am}\xi_k B^{kn}-B^{ak}\xi_k B^{mn}\bigr)\nabla_{\!m}\nabla_{\!n}Z\\
    &\mspace{103mu}-\beta\bigl(\nabla_{\!m}(B^{am}\xi_k B^{kn}-B^{ak}\xi_k B^{mn})\bigr)\nabla_{\!n}Z\;.
 \end{split}
\end{equation}
In the last step we used the Ricci identities to interchange covariant derivatives, producing a curvature term which canceled the term with Ricci tensor. Here we also used that $B^a{}_b$ commutes with $R^a{}_b$ as matrices, since $h^a{}_b$ commutes with $R^a{}_b$. Next we used twice the expression \eqref{Bder}. Pushing $\xi_k B^{kn}$ and $B^{mn}$ in the second term in the last expression back under the derivative gives
\begin{align}
 \begin{split}
    - \nabla_{\!m}&\nabla^m (B^{an}\nabla_{\!n}Z) + R^a{}_m B^{mn}\nabla_{\!n}Z\\
    =&-B^{an}\nabla_{\!n}\Box Z
    +2\beta B^{ak}\xi_k \nabla_{\!m}(B^{mn}\nabla_{\!n}Z)
      -2\beta B^{am}\nabla_{\!m}\bigl(\xi_k B^{kn}\nabla_{\!n}Z\bigr)\\
    &+\beta\Bigl(B^{am}\nabla_{\!m}(\xi_k B^{kn})
      -\xi_k B^{kn}\nabla_{\!m}B^{am}
      +B^{mn}\nabla_{\!m}(B^{ak}\xi_k)
      -B^{ak}\xi_k\nabla_{\!m}B^{mn}\Bigr)\nabla_{\!n}Z
 \end{split}\notag
\displaybreak[0]\\
 \begin{split}
    =&-B^{an}\nabla_{\!n}\Box Z
    +2\beta B^{ak}\xi_k \nabla_{\!m}(B^{mn}\nabla_{\!n}Z)
    -2\beta B^{am}\nabla_{\!m}\bigl(\xi_k B^{kn}\nabla_{\!n}Z\bigr)\\
    &+\beta\Bigl(B^{am}(\nabla_{\!m}\xi_k) B^{kn}+B^{am}\xi_k(\nabla_{\!m} B^{kn})
      \frac{\beta}{A}\bigl(k^{am}\xi_m-k^c{}_c B^{am}\xi_m\bigr)\xi_k B^{kn}\\
    &\quad+  B^{am}(\nabla_{\!k}\xi_m) B^{kn}+(\nabla_{\!m} B^{ak})\xi_k B^{mn}
      -\frac{\beta}{A} B^{ak}\xi_k\bigl(k^c{}_c\xi_m B^{mn}-\xi_m k^{mn}\bigr)
    \Bigr)\nabla_{\!n}Z\;,
 \end{split}\raisetag{12ex}\label{boxA2}
\end{align}
where we used relations \eqref{Bdiv}. Because $\ts{\xi}$ is a Killing vector, terms with $\covd\ts{\xi}$ cancel each other, as well as terms proportional to $k$. Using once more \eqref{Bder}, we obtain
\begin{align}
    - \nabla_{\!m}&\nabla^m (B^{an}\nabla_{\!n}Z) + R^a{}_m B^{mn}\nabla_{\!n}Z\notag\\
\begin{split}
    =&-B^{an}\nabla_{\!n}\Box Z
    +2\beta B^{ak}\xi_k \nabla_{\!m}(B^{mn}\nabla_{\!n}Z)
    -2\beta B^{am}\nabla_{\!m}\bigl(\xi_k B^{kn}\nabla_{\!n}Z\bigr)\\
    &+\beta^2\Bigl(-\frac{1}{A}
     \bigl(k^{am}\xi_m \xi_k B^{kn}- B^{ak}\xi_k \xi_m k^{mn}\bigr)
     +B^{am}\xi_k\bigl(B^k{}_{m}\xi_l B^{ln}-B^{kl}\xi_l B_m{}^n\bigr)\\
    &\mspace{300mu}+\bigl(B^a{}_{m}\xi_l B^{lk}-B^{al}\xi_l B_m{}^k\bigr)\xi_k B^{mn}
    \Bigr)\nabla_{\!n}Z
\end{split}
\displaybreak[0]\notag\\
\begin{split}
    =&-B^{an}\nabla_{\!n}\Box Z
    +2\beta B^{ak}\xi_k \nabla_{\!m}(B^{mn}\nabla_{\!n}Z)
    -2\beta B^{am}\nabla_{\!m}\bigl(\xi_k B^{kn}\nabla_{\!n}Z\bigr)\\
    &+\frac{\beta^2}{A}
      \Bigl(-k^{al}\xi_l \xi_k B^{kn}+ B^{ak}\xi_k \xi_l k^{ln}
      + k^{ak}\xi_k \xi_l B^{ln}-B^{al}\xi_l \xi_k k^{kn}
    \Bigr)\nabla_{\!n}Z
\end{split}\label{boxA3}\\
    =&-B^{an}\nabla_{\!n}\Box Z
    +2\beta B^{ak}\xi_k \nabla_{\!m}(B^{mn}\nabla_{\!n}Z)
    -2\beta B^{am}\nabla_{\!m}\bigl(\xi_k B^{kn}\nabla_{\!n}Z\bigr)
    \;,\notag
\end{align}
where we canceled the terms proportional to $\xi_k\xi_l B^{kl}$ and used relation \eqref{BBisAK}.


\begin{thebibliography}{10}

\bibitem{Carter:1968cmp}
B.~Carter, \emph{{Hamilton-Jacobi} and {S}chrodinger separable solutions of
  {E}instein's equations}, {\emph{Commun. Math. Phys.} {\bfseries 10} (1968)
  280--310}.

\bibitem{Teukolsky:1972}
S.~A. Teukolsky, \emph{Rotating black holes - separable wave equations for
  gravitational and electromagnetic perturbations},
  \href{https://doi.org/10.1103/PhysRevLett.29.1114}{\emph{Phys. Rev. Lett.}
  {\bfseries 29} (1972) 1114--1118}.

\bibitem{Teukolsky:1973}
S.~A. Teukolsky, \emph{Perturbations of a rotating black hole. {I}.
  {F}undamental equations for gravitational electromagnetic and neutrino field
  perturbations}, \href{https://doi.org/10.1086/152444}{\emph{Astrophys. J.}
  {\bfseries 185} (1973) 635--647}.

\bibitem{Unruh:1973}
W.~G. Unruh, \emph{Separability of the neutrino equations in a {K}err
  background}, {\emph{Phys. Rev. Lett.} {\bfseries 31} (1973) 1265}.

\bibitem{Chandrasekhar:1976}
S.~Chandrasekhar, \emph{The solution of {D}irac's equation in {K}err geometry},
  {\emph{Proc. R. Soc. Lond., Ser A} {\bfseries 349} (1976) 571--575}.

\bibitem{Page:1976}
D.~N. Page, \emph{Dirac equation around a charged, rotating black hole},
  \href{https://doi.org/10.1103/PhysRevD.14.1509}{\emph{Phys. Rev. D}
  {\bfseries 14} (1976) 1509--1510}.

\bibitem{FrolovStojkovic:2003a}
V.~P. Frolov and D.~Stojkovi\'{c}, \emph{Quantum radiation from a 5-dimensional
  rotating black hole},
  \href{https://doi.org/10.1103/PhysRevD.67.084004}{\emph{Phys. Rev. D}
  {\bfseries 67} (2003) 084004},
  [\href{https://arxiv.org/abs/gr-qc/0211055}{{\ttfamily gr-qc/0211055}}].

\bibitem{Kunduri:2005fq}
H.~K. Kunduri and J.~Lucietti, \emph{Integrability and the {K}err-{(A)dS} black
  hole in five dimensions},
  \href{https://doi.org/10.1103/PhysRevD.71.104021}{\emph{Phys. Rev. D}
  {\bfseries 71} (2005) 104021},
  [\href{https://arxiv.org/abs/hep-th/0502124}{{\ttfamily hep-th/0502124}}].

\bibitem{Vasudevan:2005js}
M.~Vasudevan and K.~A. Stevens, \emph{Integrability of particle motion and
  scalar field propagation in {K}err-(anti) {de Sitter} black hole spacetimes
  in all dimensions},
  \href{https://doi.org/10.1103/PhysRevD.72.124008}{\emph{Phys. Rev. D}
  {\bfseries 72} (2005) 124008},
  [\href{https://arxiv.org/abs/gr-qc/0507096}{{\ttfamily gr-qc/0507096}}].

\bibitem{Vasudevan:2004mr}
M.~Vasudevan, K.~A. Stevens and D.~N. Page, \emph{Particle motion and scalar
  field propagation in {M}yers-{P}erry black hole spacetimes in all
  dimensions}, \href{https://doi.org/10.1088/0264-9381/22/7/017}{\emph{Class.
  Quantum Grav.} {\bfseries 22} (2005) 1469--1482},
  [\href{https://arxiv.org/abs/gr-qc/0407030}{{\ttfamily gr-qc/0407030}}].

\bibitem{Vasudevan:2004ca}
M.~Vasudevan, K.~A. Stevens and D.~N. Page, \emph{Separability of the
  {H}amilton-{J}acobi and {K}lein-{G}ordon equations in {K}err-{de Sitter}
  metrics}, \href{https://doi.org/10.1088/0264-9381/22/2/007}{\emph{Class.
  Quantum Grav.} {\bfseries 22} (2005) 339--352},
  [\href{https://arxiv.org/abs/gr-qc/0405125}{{\ttfamily gr-qc/0405125}}].

\bibitem{Davis:2006hy}
P.~Davis, \emph{A killing tensor for higher dimensional {K}err-{AdS} black
  holes with {NUT} charge},
  \href{https://doi.org/10.1088/0264-9381/23/10/023}{\emph{Class. Quantum
  Grav.} {\bfseries 23} (2006) 3607--3618},
  [\href{https://arxiv.org/abs/hep-th/0602118}{{\ttfamily hep-th/0602118}}].

\bibitem{Chen:2006ui}
W.~Chen, H.~Lu and C.~N. Pope, \emph{Separability in cohomogeneity-2
  {K}err-{NUT}-{AdS} metrics},
  \href{https://doi.org/10.1088/1126-6708/2006/04/008}{\emph{JHEP} {\bfseries
  0604} (2006) 008}, [\href{https://arxiv.org/abs/hep-th/0602084}{{\ttfamily
  hep-th/0602084}}].

\bibitem{Kubiznak:2006kt}
D.~Kubiz\v{n}\'{a}k and V.~P. Frolov, \emph{Hidden symmetry of higher
  dimensional {K}err-{NUT}-{AdS} spacetimes},
  \href{https://doi.org/10.1088/0264-9381/24/3/F01}{\emph{Class. Quantum Grav.}
  {\bfseries 24} (2007) F1--F6},
  [\href{https://arxiv.org/abs/gr-qc/0610144}{{\ttfamily gr-qc/0610144}}].

\bibitem{Chen:2006xh}
W.~Chen, H.~Lu and C.~N. Pope, \emph{General {K}err-{NUT}-{AdS} metrics in all
  dimensions}, \href{https://doi.org/10.1088/0264-9381/23/17/013}{\emph{Class.
  Quantum Grav.} {\bfseries 23} (2006) 5323--5340},
  [\href{https://arxiv.org/abs/hep-th/0604125}{{\ttfamily hep-th/0604125}}].

\bibitem{Frolov:2006pe}
V.~P. Frolov, P.~Krtou\v{s} and D.~Kubiz\v{n}\'{a}k, \emph{Separability of
  {Hamilton-Jacobi} and {Klein-Gordon} equations in general {K}err-{NUT}-{AdS}
  spacetimes}, \href{https://doi.org/10.1088/1126-6708/2007/02/005}{\emph{JHEP}
  {\bfseries 0702} (2007) 005},
  [\href{https://arxiv.org/abs/hep-th/0611245}{{\ttfamily hep-th/0611245}}].

\bibitem{FrolovKrtousKubiznak:2017review}
V.~P. Frolov, P.~Krtou\v{s} and D.~Kubiz\v{n}\'ak, \emph{Black holes, hidden
  symmetries, and complete integrability},
  \href{https://doi.org/10.1007/s41114-017-0009-9}{\emph{Living Rev. Rel.}
  {\bfseries 20} (2017) 6}, [\href{https://arxiv.org/abs/1705.05482}{{\ttfamily
  1705.05482}}].

\bibitem{carter1987separability}
B.~Carter, \emph{Separability of the {K}illing--{M}axwell system underlying the
  generalized angular momentum constant in the {K}err--{N}ewman black hole
  metrics}, {\emph{J. Math. Phys.} {\bfseries 28} (1987) 1535--1538}.

\bibitem{Kolar:2015cha}
I.~Kol\'{a}\v{r} and P.~Krtou\v{s}, \emph{Weak electromagnetic field admitting
  integrability in {K}err-{NUT}-({A})ds spacetimes},
  \href{https://doi.org/10.1103/PhysRevD.91.124045}{\emph{Phys. Rev. D}
  {\bfseries 91} (2015) 124045},
  [\href{https://arxiv.org/abs/1504.00524}{{\ttfamily 1504.00524}}].

\bibitem{Sergyeyev:2007gf}
A.~Sergyeyev and P.~Krtou\v{s}, \emph{Complete set of commuting symmetry
  operators for {K}lein-{G}ordon equation in generalized higher-dimensional
  {K}err-{NUT}-{(A)dS} spacetimes},
  \href{https://doi.org/10.1103/PhysRevD.77.044033}{\emph{Phys. Rev. D}
  {\bfseries 77} (2008) 044033},
  [\href{https://arxiv.org/abs/0711.4623}{{\ttfamily 0711.4623}}].

\bibitem{Frolov:2010cr}
V.~P. Frolov and P.~Krtou\v{s}, \emph{Charged particle in higher dimensional
  weakly charged rotating black hole spacetime},
  \href{https://doi.org/10.1103/PhysRevD.83.024016}{\emph{Phys. Rev. D}
  {\bfseries 83} (2011) 024016},
  [\href{https://arxiv.org/abs/1010.2266}{{\ttfamily 1010.2266}}].

\bibitem{Houri:2007xz}
T.~Houri, T.~Oota and Y.~Yasui, \emph{Closed conformal {K}illing-{Y}ano tensor
  and {K}err-{NUT}-{de Sitter} spacetime uniqueness},
  \href{https://doi.org/10.1016/j.physletb.2007.09.034}{\emph{Phys. Lett. B}
  {\bfseries 656} (2007) 214--216},
  [\href{https://arxiv.org/abs/0708.1368}{{\ttfamily 0708.1368}}].

\bibitem{Krtous:2008tb}
P.~Krtou\v{s}, V.~P. Frolov and D.~Kubiz\v{n}\'{a}k, \emph{Hidden symmetries of
  higher dimensional black holes and uniqueness of the {K}err-{NUT}-{(A)dS}
  spacetime}, \href{https://doi.org/10.1103/PhysRevD.78.064022}{\emph{Phys.
  Rev. D} {\bfseries 78} (2008) 064022},
  [\href{https://arxiv.org/abs/0804.4705}{{\ttfamily 0804.4705}}].

\bibitem{OotaYasui:2008}
T.~Oota and Y.~Yasui, \emph{Separability of {D}irac equation in higher
  dimensional {K}err-{NUT}-{de Sitter} spacetime},
  \href{https://doi.org/10.1016/j.physletb.2007.11.057}{\emph{Phys. Lett. B}
  {\bfseries 659} (2008) 688--693},
  [\href{https://arxiv.org/abs/0711.0078}{{\ttfamily 0711.0078}}].

\bibitem{Cariglia:2011qb}
M.~Cariglia, P.~Krtou\v{s} and D.~Kubiz\v{n}\'{a}k, \emph{Dirac equation in
  {Kerr-NUT-(A)dS} spacetimes: Intrinsic characterization of separability in
  all dimensions},
  \href{https://doi.org/10.1103/PhysRevD.84.024008}{\emph{Phys. Rev. D}
  {\bfseries 84} (2011) 024008},
  [\href{https://arxiv.org/abs/1104.4123}{{\ttfamily 1104.4123}}].

\bibitem{Cariglia:2011yt}
M.~Cariglia, P.~Krtou\v{s} and D.~Kubiz\v{n}\'{a}k, \emph{Commuting symmetry
  operators of the {D}irac equation, {K}illing-{Y}ano and {Schouten-Nijenhuis}
  brackets}, \href{https://doi.org/10.1103/PhysRevD.84.024004}{\emph{Phys. Rev.
  D} {\bfseries 84} (2011) 024004},
  [\href{https://arxiv.org/abs/1102.4501}{{\ttfamily 1102.4501}}].

\bibitem{CarigliaEtal:2012b}
M.~Cariglia, V.~P. Frolov, P.~Krtou\v{s} and D.~Kubiz\v{n}\'{a}k,
  \emph{Electron in higher-dimensional weakly charged rotating black hole
  spacetimes}, \href{https://doi.org/10.1103/PhysRevD.87.064003}{\emph{Phys.
  Rev. D} {\bfseries 87} (2013) 064003},
  [\href{https://arxiv.org/abs/1211.4631}{{\ttfamily 1211.4631}}].

\bibitem{Kubiznak:2010ig}
D.~Kubiz\v{n}\'{a}k, C.~M. Warnick and P.~Krtou\v{s}, \emph{Hidden symmetry in
  the presence of fluxes},
  \href{https://doi.org/10.1016/j.nuclphysb.2010.11.001}{\emph{Nucl. Phys. B}
  {\bfseries 844} (2011) 185--198},
  [\href{https://arxiv.org/abs/1009.2767}{{\ttfamily 1009.2767}}].

\bibitem{Kunduri:2006qa}
H.~K. Kunduri, J.~Lucietti and H.~S. Reall, \emph{Gravitational perturbations
  of higher dimensional rotating black holes: {T}ensor perturbations},
  \href{https://doi.org/10.1103/PhysRevD.74.084021}{\emph{Phys. Rev. D}
  {\bfseries 74} (2006) 084021},
  [\href{https://arxiv.org/abs/hep-th/0606076}{{\ttfamily hep-th/0606076}}].

\bibitem{Oota:2008uj}
T.~Oota and Y.~Yasui, \emph{Separability of gravitational perturbation in
  generalized {K}err-{NUT}-{de Sitter} spacetime},
  \href{https://doi.org/10.1142/S0217751X10049001}{\emph{Int. J. Mod. Phys.}
  {\bfseries A25} (2010) 3055--3094},
  [\href{https://arxiv.org/abs/0812.1623}{{\ttfamily 0812.1623}}].

\bibitem{Lunin:2017}
O.~Lunin, \emph{Maxwell's equations in the {Myers-Perry} geometry},
  \href{https://doi.org/10.1007/JHEP12(2017)138}{\emph{JHEP} {\bfseries 1712}
  (2017) 138}, [\href{https://arxiv.org/abs/1708.06766}{{\ttfamily
  1708.06766}}].

\bibitem{Frolov:2008jr}
V.~P. Frolov and D.~Kubiz\v{n}\'{a}k, \emph{Higher-dimensional black holes:
  Hidden symmetries and separation of variables},
  \href{https://doi.org/10.1088/0264-9381/25/15/154005}{\emph{Class. Quantum
  Grav.} {\bfseries 25} (2008) 154005},
  [\href{https://arxiv.org/abs/0802.0322}{{\ttfamily 0802.0322}}].

\bibitem{Chervonyi:2015ima}
Y.~Chervonyi and O.~Lunin, \emph{Killing(-{Y}ano) tensors in string theory},
  \href{https://doi.org/10.1007/JHEP09(2015)182}{\emph{JHEP} {\bfseries 1509}
  (2015) 182}, [\href{https://arxiv.org/abs/1505.06154}{{\ttfamily
  1505.06154}}].

\bibitem{Krtous:2007}
P.~Krtou\v{s}, \emph{Electromagnetic field in higher-dimensional black-hole
  spacetimes}, \href{https://doi.org/10.1103/PhysRevD.76.084035}{\emph{Phys.
  Rev. D} {\bfseries 76} (2007) 084035},
  [\href{https://arxiv.org/abs/0707.0002}{{\ttfamily 0707.0002}}].

\bibitem{Chen:2007fs}
W.~Chen and H.~Lu, \emph{{Kerr-Schild} structure and harmonic 2-forms on
  {(A)dS-Kerr-NUT} metrics},
  \href{https://doi.org/10.1016/j.physletb.2007.09.066}{\emph{Phys. Lett. B}
  {\bfseries 658} (2008) 158--163},
  [\href{https://arxiv.org/abs/0705.4471}{{\ttfamily 0705.4471}}].

\bibitem{MyersPerry:1986}
R.~C. Myers and M.~J. Perry, \emph{Black holes in higher dimensional
  space-times}, \href{https://doi.org/10.1016/0003-4916(86)90186-7}{\emph{Ann.
  Phys. (N.Y.)} {\bfseries 172} (1986) 304--347}.

\bibitem{Gibbons:2004uw}
G.~W. Gibbons, H.~Lu, D.~N. Page and C.~N. Pope, \emph{The general {K}err-{de
  Sitter} metrics in all dimensions},
  \href{https://doi.org/10.1016/j.geomphys.2004.05.001}{\emph{J. Geom. Phys.}
  {\bfseries 53} (2005) 49--73},
  [\href{https://arxiv.org/abs/hep-th/0404008}{{\ttfamily hep-th/0404008}}].

\bibitem{Gibbons:2004js}
G.~W. Gibbons, H.~Lu, D.~N. Page and C.~N. Pope, \emph{Rotating black holes in
  higher dimensions with a cosmological constant},
  \href{https://doi.org/10.1103/PhysRevLett.93.171102}{\emph{Phys. Rev. Lett.}
  {\bfseries 93} (2004) 171102},
  [\href{https://arxiv.org/abs/hep-th/0409155}{{\ttfamily hep-th/0409155}}].

\bibitem{Krtous:2006qy}
P.~Krtou\v{s}, D.~Kubiz\v{n}\'{a}k, D.~N. Page and V.~P. Frolov,
  \emph{{K}illing-{Y}ano tensors, rank-2 {K}illing tensors, and conserved
  quantities in higher dimensions},
  \href{https://doi.org/10.1088/1126-6708/2007/02/004}{\emph{JHEP} {\bfseries
  0702} (2007) 004}, [\href{https://arxiv.org/abs/hep-th/0612029}{{\ttfamily
  hep-th/0612029}}].

\bibitem{FrolovKrtousKubiznak:2018a}
V.~P. Frolov, P.~Krtou\v{s} and D.~Kubiz\v{n}\'ak, \emph{Separation variables
  in {M}axwell equations in {Pleba\'{n}ski--Demia\'{n}ski metric}},
  \href{https://doi.org/10.1103/PhysRevD.97.101701}{\emph{Phys. Rev. D}
  {\bfseries 97} (2018) 101701(R)},
  [\href{https://arxiv.org/abs/1802.09491}{{\ttfamily 1802.09491}}].

\bibitem{Proca:1936}
A.~Proca, \emph{Sur la th\'{e}orie ondulatoire des \'{e}lectrons positifs et
  n\'{e}gatifs}, {\emph{Journal de Physique et le Radium} {\bfseries 7} (1936)
  347--353}.

\bibitem{Belinfante:1949}
F.~J. Belinfante, \emph{The interaction representation of the {P}roca field},
  \href{https://doi.org/10.1103/PhysRev.76.66}{\emph{Phys. Rev.} {\bfseries 76}
  (1949) 66--80}.

\bibitem{Rosen:1994}
N.~Rosen, \emph{A classical {P}roca particle}, {\emph{Found. Phys.} {\bfseries
  24} (12, 1994) 1689--1695}.

\bibitem{Seitz:1986sc}
M.~Seitz, \emph{Proca field in a space-time with curvature and torsion},
  \href{https://doi.org/10.1088/0264-9381/3/6/023}{\emph{Class. Quantum Grav.}
  {\bfseries 3} (1986) 1265--1273}.

\bibitem{Frolov:2018ezx}
V.~P. Frolov, P.~Krtou\v{s}, D.~Kubiz\v{n}\'ak and J.~E. Santos, \emph{Massive
  vector fields in rotating black-hole spacetimes: {S}eparability and
  quasinormal modes},
  \href{https://doi.org/10.1103/PhysRevLett.120.231103}{\emph{Phys. Rev. Lett.}
  {\bfseries 120} (2018) 231103},
  [\href{https://arxiv.org/abs/1804.00030}{{\ttfamily 1804.00030}}].

\bibitem{Chong:2004hw}
Z.~W. Chong, G.~W. Gibbons, H.~Lu and C.~N. Pope, \emph{Separability and
  {K}illing tensors in {Kerr-Taub-NUT-de~Sitter} metrics in higher dimensions},
  \href{https://doi.org/10.1016/j.physletb.2004.07.066}{\emph{Phys. Lett. B}
  {\bfseries 609} (2005) 124--132},
  [\href{https://arxiv.org/abs/hep-th/0405061}{{\ttfamily hep-th/0405061}}].

\bibitem{Kolar:2017vjl}
I.~Kol\'a\v{r} and P.~Krtou\v{s}, \emph{{NUT}-like and near-horizon limits of
  {Kerr-NUT-(A)dS} spacetimes},
  \href{https://doi.org/10.1103/PhysRevD.95.124044}{\emph{Phys. Rev. D}
  {\bfseries D95} (2017) 124044},
  [\href{https://arxiv.org/abs/1701.03950}{{\ttfamily 1701.03950}}].

\bibitem{Krtous:2015zco}
P.~Krtou\v{s}, D.~Kubiz\v{n}\'{a}k, V.~P. Frolov and I.~Kol\'{a}\v{r},
  \emph{Deformed and twisted black holes with {NUT}s},
  \href{https://doi.org/10.1088/0264-9381/33/11/115016}{\emph{Class. Quantum
  Grav.} {\bfseries 33} (2016) 115016},
  [\href{https://arxiv.org/abs/1511.02536}{{\ttfamily 1511.02536}}].

\bibitem{KolarKrtous:2018}
I.~Kol\'a\v{r} and P.~Krtou\v{s}, ``Fixed points of isometries in
  {Kerr-NUT-(A)dS} spacetimes.'' 2018.

\end{thebibliography}

\providecommand{\href}[2]{#2}\begingroup\raggedright\endgroup

\end{document}